\documentclass[journal=jpcbfk,manuscript=article]{achemso}
\usepackage[utf8]{inputenc}

\usepackage{longtable}
\usepackage{multirow}
\usepackage{amsmath}
\usepackage[usenames,dvipsnames]{color}
\usepackage{soul}
\usepackage{threeparttable}
\usepackage{graphicx}
\usepackage{amsmath,amssymb}
\usepackage{caption}
\usepackage{color}
\usepackage{xcolor}
\usepackage{dcolumn}
\usepackage{bm}
\usepackage{float}
\usepackage{subcaption}
\usepackage{textcomp}
\usepackage{url}
\usepackage[normalem]{ulem}

\usepackage{microtype}
\usepackage{etoolbox}

\usepackage{xr}
\makeatletter
\newcommand*{\addFileDependency}[1]{
  \typeout{(#1)}
  \@addtofilelist{#1}
  \IfFileExists{#1}{}{\typeout{No file #1.}}
}
\makeatother
\newcommand*{\myexternaldocument}[1]{
    \externaldocument{#1}
    \addFileDependency{#1.tex}
    \addFileDependency{#1.aux}
}
\myexternaldocument{si}
\usepackage[unicode=true,
            bookmarks=true,
            bookmarksnumbered=true,
            bookmarksopen=true,
            bookmarksopenlevel=2,
            breaklinks=false,
            pdfborder={0 0 1},
            backref=false,
            colorlinks=true,
            hidelinks
]{hyperref}
\author{Juan Carlos San Vicente Veliz} \affiliation{Department of
  Chemistry, University of Basel, Klingelbergstrasse 80, CH-4056
  Basel, Switzerland} \altaffiliation{Department of Chemistry, Temple
  University, Philadelphia, Pennsylvania 19122, United
  States}\altaffiliation{These authors contributed equally}

\author{Sung Min Jo} \affiliation{Department of Aerospace Engineering,
  Korea Advanced Institute of Science and Technology, Daejeon, 34141,
  Republic of Korea} \altaffiliation{These authors contributed
  equally}
  
\author{Jingchun Wang} \affiliation{Department of Chemistry,
  University of Basel, Klingelbergstrasse 80, CH-4056 Basel,
  Switzerland}

\author{Raymond J. Bemish} \affiliation{Air Force Research Laboratory,
  Space Vehicles Directorate, Kirtland AFB, New Mexico 87117, USA}

\author{Markus Meuwly}\email{m.meuwly@unibas.ch}
\affiliation{Department of Chemistry, University of Basel,
  Klingelbergstrasse 80, CH-4056 Basel, Switzerland}
\altaffiliation{Department of Chemistry, Brown University, Providence,
  RI, USA}
 
\title[]{Reaction Dynamics for the [NNO] System from State-Resolved
  and Coarse-Grained Models}

\begin{document}
\date{\today}

\begin{abstract}
The dynamics for the NO($X^2 \Pi$) + N($^4$S) $\leftrightarrow$
N$_{2}(X^{1}\Sigma_{g}^{+}$) + O($^{3}$P) reaction was followed in the $^3$A' electronic state using state-to-state (STS) and Arrhenius-based rates from two different high-level potential energy surfaces represented as a reproducing kernel (RKHS) and permutationally invariant polynomials (PIPs). Despite the different number of bound states supported by the RKHS- and PIP-PESs the ignition points from STS and Arrhenius rates are at $\sim 10^{-6}$ s  whether or not reverse rates are from assuming microreversibility or explicitly given. Conversion from NO to N$_2$ is incomplete if Arrhenius-rates are used but complete turnover is observed if STS-information is used. This is due to non-equilibrium energy flow and state dynamics which requires a state-based description. Including full dissociation leads asymptotically to the correct 2:1 [N]:[O] concentration with little differences for the species' dynamics depending on the PES used for the STS-information. In conclusion, concentration profiles from coarse-grained simulations are consistent over 14 orders of magnitude in time using STS-information based on two different high-level PESs.
\end{abstract}

\section{Introduction}
The species and internal-state evolution of reacting chemical systems
is of central importance in combustion and hypersonics. Characterizing
the temporal evolution of the underlying chemistry is important for
describing the energy content and energy redistribution in such
energized environments. In hypersonic flight,\cite{anderson.book}
objects traveling through atmospheres at high speed ($>$ Mach 5)
dissipate large amounts of energy to the surrounding gas. This
generates highly non-equilibrium conditions with respect to occupation
of translational, rotational, vibrational, and electronic degrees of
freedom of the molecules constituting such atmospheres. At such speeds
and in particular in the bow shock region surrounding the traveling
vehicle, the energies (and hence temperatures) are sufficiently high
to dissociate small molecules such as N$_2$ and O$_2$. Earth's
atmosphere features sufficiently dense regions to feature frequent
molecular collisions (i.e. the troposphere and stratosphere) between
O$_2$, N$_2$, and NO formed in the hypersonic flow. Mars, Titan,
Venus, and other planets with dense atmospheres have significantly
more complex polyatomic species to consider.\cite{horst:2012}\\

\noindent
Hypersonic flight is an endeavor on a grand scale. Objects travel at
speeds of kilometers per second, generating bow shock with
temperatures up to 20000 K and surface temperatures only limited by
the vaporization temperature of its outer shell. At present the
current upper limit for velocity of a man-made object is 12.5 km/s
(Stardust capsule\cite{qualls:2010}) although natural meteors can
reach considerably higher velocities ($\sim 70$ km/s) as reported for
Meteor Leonid.\cite{boyd:2000} At such high speeds the chemistry and
thermodynamics of material flow are coupled. For Earth's atmosphere,
the chemistry primarily involves the dissociation of diatomic
molecules (O$_2$, NO, and N$_2$) into atomic fragments which removes
thermal energy but generates chemically highly reactive
species. Chemical processes, including vibrational relaxation and
dissociation occur on the picosecond time scale compared with the
second time scale on which the object travels. Similarly, the length
scales involved span 12 orders of magnitude extending from 1 \AA\/ for
chemical bonds to 1 m or more for the object's size. In other words,
hypersonics is inherently a multi-scale problem.\\

\noindent
In addition to length and time scales, the chemistry under such
circumstances also covers wide ranges in internal state space of the
molecules involved. To illustrate this it is noted that a diatomic
molecule such as NO or N$_2$ has of the order of $10^4$ available
rovibrational states $[v,j]$. A reaction A+BC$\rightarrow$B+AC
therefore leads to approximately $10^8$ state-to-state (STS) cross
sections. To converge each of those using quasiclassical trajectory
(QCT)-type simulations requires $\sim 10^5$ trajectory simulations for
one value of the collision energy. Hence, a converged set of STS cross
sections for one reaction would involve $\sim 10^{14}$ QCT
simulations. Modeling the nonequilibrium chemical reaction dynamics on
the relevant temporal and spatial scales is extremely challenging and
using explicit QCT-based simulations in meso- to macroscopic
simulations for characterizing the reactive flow around bodies moving
at hypersonic speed is clearly unfeasible.\\

\noindent
For this reason more coarse-grained approaches rooted in computational
fluid dynamics (CFD) have been
developed.\cite{munafo2014boltzmann,kim2021modification,macdonald2018construction,chaudhry2020implementation,sahai2017adaptive,liu2015general,Venturi2020}
One such approach employs systems of coupled rate equations that can
be solved using standard linear algebra tools. The ingredients
required are the STS cross sections for all bimolecular processes
including dissociation of all species. Among others, such master
equation analyses have been carried out for the [NNO] system in order
to characterize the microscopic behavior of the ro-vibrational states,
and corresponding macroscopic energy and concentration
traces.\cite{panesi:2022} Along similar lines, the combined [NNO] and
[NOO] systems were studied in the hypersonic regime to identify their
relative contributions in the context of microscopic ro-vibrational
energy transfers.\cite{rodriguez:2025} By using a Direct Molecular
Simulation (DMS) method,\cite{torres2024high} the average amount of
ro-vibrational energy change during the Zeldovich
process\cite{bose1996kinetics,luo2017ab,panesi:2022,rodriguez:2025,torres2024high,valentini2024toward}
was quantified.\cite{torres2024high} This DMS study was further
extended to the 5-species air system (N, O, N$_2$, O$_2$, NO) to
simulate two-dimensional axisymmetric hypersonic
flows.\cite{valentini2024toward}\\

\noindent
Master equation- and DMS-based studies use collision-dynamics
simulations on a reactive potential energy surface (PES) as a starting
point. The Master equation approach requires a dictionary of STS cross
sections which can be built from different approaches. Here, the
coarse-grained and state-resolved chemical dynamics for the [NNO]
system, focusing on the NO($X^2 \Pi$) + N($^4$S) $\leftrightarrow$
N$_{2}(X^{1}\Sigma_{g}^{+}$) + O($^{3}$P) reaction, in a particular
electronic state ($^3$A$'$). For this, two high-level PESs are
available. The first, referred to as PES$_{\rm B}$ (Basel PES), was
based on MRCI+Q calculations and represented as a reproducing kernel
Hilbert space (RKHS)\cite{MM.n2o:2020} whereas the second, PES$_{\rm
  M}$ (Minnesota PES), based on a MRCI+Q level calculation, was fit to
permutationally invariant polynomials (PIP).\cite{lin:2016} PES$_{\rm
  B}$ was used successfully to calculate thermal rates between 100 and
20000 K for the downhill reaction NO(X$^2 \Pi$) + N($^4$S)
$\rightarrow$ N$_2$(X$^1 \Sigma_{\rm g}^{+}$) +
O($^{3}$P).\cite{MM.n2o:2020} Simulations for the reverse, uphill
reaction were carried out between 3000 and 20000 K. Additionally,
vibrational relaxation was explored for O + N$_{2}(\nu =1) \rightarrow
$ O + N$_{2}(\nu' =0)$ at temperatures between 1000 and 10000
K.\cite{MM.n2o:2020} PES$_{\rm M}$ was used for studying the thermal
rates for the endothermic forward reaction N$_{2}$ + O $\rightarrow$
NO + N, on both, the $^{3}$A$'$ and $^{3}$A$''$
PESs.\cite{panesi:2022}\\

\noindent
One of the central aspects of simulation studies involving multiple
spatial and temporal scales is the propagation of uncertainties. For
the type of study carried out in the present work, uncertainties can
arise from a) the level of quantum chemical theory at which the
underlying PESs were determined, b) the procedure used to represent
the PESs for use in nuclear motion simulation studies, c) how the
state-to-state information was determined, and d) whether quantum
mechanical or classical dynamics methods were used. In the context of
this study, points a) to c) are considered by using two different
approaches for calculating and representing the PESs, and for
obtaining the necessary STS cross sections. The coarse grained
simulations were then carried out using the PLATO (PLAsmas in
Thermodynamic nOn-equilibrium) software.\cite{munafo:2020} Point d)
has, for example, been investigated for the C+O$_2$ or HeH$^+$ + H
reactions for which QCT and TIQM simulations yield comparable reaction
rates.\cite{MM.co2quantum:2022,MM.heh:2025} Earlier work on three-body
collisions for the O+O$_2$ and N+N$_2$ systems provided some
uncertainty quantification and highlight potential limitations of
QCT-based studies.\cite{geistfeld:2023} Also, in the context of
computational fluid dynamics simulations, uncertainty quantification
and sensitivity analysis were carried out for hypersonic
flows.\cite{lockwood:2013,west:2015,holloway:2022} Such analyses for
example found, that the O$_2$+O$_2$ reaction is less important than
those involving N$_2$.\cite{holloway:2022} \\

\noindent
The main aim of the present work is to determine the time evolution of
the chemical composition of the reactive [N, O, N$_{2}$, NO] system
using two different PESs, describing the reaction dynamics through
Arrhenius rates (coarse grained) and STS cross sections
(state-resolved) using neural network-(NN) based and explicitly
calculated STS-information from QCT simulations. First, the methods
used are introduced. This is followed by results from the
coarse-grained simulations, which include a test of the
microreversibility (MR) assumption by comparing with simulations that
use explicitly calculated reverse rates. Next, the impact of using
Arrhenius versus STS rates in the Master equations on the
time-dependent species concentrations is examined. Finally, the
influence of the chosen PES on the chemical network predictions is
assessed, and key conclusions are drawn.\\

\section{Methods}

\subsection{The Potential Energy Surfaces}
The two PESs for the $^3$A$'$ state of the [NNO] system used in the
present work were based on MRCI+Q calculations with the (m)aug-cc-pVTZ
basis sets.\cite{MM.n2o:2020,lin:2016} Surface PES$_{\rm B}$ (for the
Basel PES)\cite{MM.n2o:2020} was represented as a reproducing kernel
Hilbert space\cite{rabitz:1996,MM.rkhs:2017,MM.rkhs:2020} whereas
PES$_{\rm M}$ (for the Minnesota PES)\cite{lin:2016} was fit to
permutationally invariant polynomials (PIPs). For the CASSCF
calculations, 10 active electrons in 9 orbitals CAS(10,9) were
used. For PES$_{\rm B}$ the CASSCF wavefunction was used for the MRCI
+ Q calculations, while a state-averaged-CASSCF calculation was
carried out prior to the MRCI calculation for PES$_{\rm M}$. Also,
PES$_{\rm M}$ applied a scaling method, mainly to better describe
dissociation of the diatomics when approaching full dissociation of
the system.\cite{lin:2016} Hence, the two PESs considered in the
present work are based on different quantum chemical calculations and
use different strategies to represent them.\\

\noindent
The two representation schemes (RKHS and PIP) both yield highly
accurate representations and allow for the computation of analytical
forces required for running MD and QCT simulations. For PES$_{\rm B}$,
a grid of 30 (O + N$_{2}$) and 28 (N + NO) grid points for the radial
coordinate $R$ between [1.4-12.0] a$_{0}$ was used with the diatomic
distance $r$ covered by 20 and 21 points, respectively, between
[1.55/1.50-4.0] a$_{0}$. Additionally, 13 Legendre quadrature angles
were used to effectively sample the angular component, yielding 7800
and 7644 grid points for the (O + N$_{2}$) and (N + NO) channels. The
$^{3}$A$'$ surface for PES$_{\rm M}$ was constructed from a
considerably smaller number of energy evaluations: 592 and 1706 grid
points for the two channels at the SA-CASSCF/MRCI level of theory. For
the two radial coordinates, the grid covered ranges [1.7--4.9] a$_0$
and [1.7--7.5] a$_{0}$ for $r$ and $R$, respectively.\\

\subsection{QCT simulations}
Quasi-Classical trajectory simulations were used to obtain
state-to-state cross sections. For PES$_{\rm B}$ specifics were
described previously\cite{MM.n2o:2020,MM.co2:2021} and hence only a
concise summary is given here together with details of new simulations
that were needed to increase the fidelity of the model. For the
thermal rates, initial reactant states ($v,j$) were sampled from a
Boltzmann distribution at 10000 K. The initial rovibrational quantum
states $(v,j)$ for NO and N$_2$, using PES$_{\rm B}$, were obtained
from the semiclassical theory of bound states.\cite{bernstein_1979}
The internal energy of the diatom, $\varepsilon_{v,j}$, is given by
the condition
\begin{equation}
  \label{eq:diat-vj}
    \frac{1}{2} J_v - \pi \hbar (v+\frac{1}{2}) = 0
\end{equation}
where
\begin{equation}
    J_v = 2\int_{q_-}^{q_+} \left\{2\mu_r
    \left[\varepsilon_{v,j}-V(r)-\frac{J_r}{2\mu_r r^2}\right]
    \right\}^{1/2} dr
\end{equation}
Using PES$_{\rm B}$, Eq.~\ref{eq:diat-vj} was solved numerically using
the Newton-Raphson method. With an in-house QCT program the 6329
initial $(v,j)$ states for NO, and 8733 for N$_2$ were determined up
to the corresponding dissociation limits.\cite{koner:2016} The $v-$
and $j-$quantum numbers reach maximum values of NO($v_{\rm max} = 47$,
$j_{\rm max} = 241)$ and N$_{2}$($v_{\rm max} = 57$, $j_{\rm max} =
273)$ for PES$_{\rm B}$.\cite{MM.sts:2019} For both the N + NO and the
O + N$_2$ reactions a total of $N_{\rm tot} = 160000$ trajectories
were run for each initial condition. The impact parameter was sampled
using stratified sampling and divided into six strata $b_{i}=[0,
  b_{\rm max}]$ with $b_{\rm max} = 14.0$ a$_0$. Final rotational and
vibrational states of the product diatom were assigned using a
Gaussian binning scheme,\cite{bonnet:2004} which results in faster
convergence.  The state-to-state cross section was then computed from
$\sigma_{\rm x}=\pi b_{\rm max}^{2} P_{\rm x}$, where
$P_{\rm x} = \frac{N_{\rm x}}{N_{\rm tot}}$ is the ratio of
trajectories that reacted for a particular initial condition.\\

\noindent
For PES$_{\rm M}$, all necessary STS information had been determined
in previous work using the
\textsc{CoarseAIR}\cite{Venturi2020_ML,Venturi2020} code which is an
updated version of the original \textsc{VVTC} code developed at NASA
Ames Research Center.\cite{SCHWENKE_VVTC_1988} These simulations also
employ stratified sampling over the impact parameter $b_i$ with
binning of 0.25 a$_0$. The maximum simulation time was $4 \times 10^5$
au with an initial time step of 5 au in the adaptive-time step
Adams-Bashforth-Moulton integrator\cite{bulirsch:1966} for each
trajectory. For the thermal rates, $10^6$ trajectories were simulated
for the initially sampled reactant states out of the Boltzmann
distribution. The maximum internal quantum numbers for PES$_{\rm M}$
are NO($v_{\rm max} = 45$, $j_{\rm max} = 251)$ and N$_{2}$($v_{\rm
  max} = 53$, $j_{\rm max} = 280)$ for PES$_{\rm M}$.\cite{lin:2016}
For the STS rates, QCT simulations were performed for both, the NO+N
and N$_2$+O reactions, using 50000 trajectories for each initial
condition. It should be noted that this number of trajectories does
not suffice to converge all state-to-state cross sections.\\

\subsection{State-to-State Information from PES$_{\rm B}$ and PES$_{\rm M}$}
For PES$_{\rm B}$, a neural network-based model to predict the full
set of STS cross sections was used. Evaluation and analysis of the
existing STS2019 model\cite{MM.sts:2019} for the NO($X^2 \Pi$) +
N($^4$S) (forward, downhill) reaction indicated that extensions with
respect to the state-space covered were required. For the highest
$(v,j)-$state of NO, the predicted cross sections did not match those
from explicit QCT simulations, as discussed further below. As a
consequence, STS2019 needed to be improved and retrained.\\

\noindent
The grid of initial conditions (IC) on which QCT simulations were
carried out for STS2019 consisted of $v \in [0, 3, 6, 9, 12, 15, 19,
  23, 28, 34]$, 12 rotational states $j \in [0, 25, 50, 75, 100, 125,
  145, 160, 175,\\ 190, 200, 210]$, and 15 collision energies $E_{t}
\in [0.05, 0.1, 0.25, 0.5, 0.8, 1.2, 1.6, 2.0, 2.5, 3.0, 3.5,\\ 4.0,
  4.5, 5.0, 5.5]$ eV which amounts to 1800 different ICs. This was
extended by running additional QCT simulations for the $v=40,46$ and
identical $j-$ and $E_{\rm t}-$ grids as before to improve the ST2019
model. For these 360 ICs, $1.6 \times 10^5$ QCT simulations were
carried out each to determine the necessary STS-information.\\

\noindent
For the reverse ("uphill"), i.e. considerably lower probability
reaction N$_{2}(X^{1}\Sigma_{g}^{+}$) + O($^{3}$P) $\rightarrow$
NO($X^2 \Pi$) + N($^4$S) a new STS model was also trained from
reference QCT simulations using PES$_{\rm B}$. For the QCT
simulations, initial conditions were generated from a predefined grid:
$v_{\rm N_2}=[0, 3, 6, 9, 12, 19, 23, 28, 34, 41, 47, 56]$; $j_{\rm
  N_2}=[0, 25, 50, 75, 100, 125, 145, 160, 175,\\ 190, 200, 210]$ and
$E_{\rm col}= [0.8, 1.2, 1.6, 2.0, 2.5, 3.0, 3.5, 4.0, 4.5, 5.5, 6.5,
  7.0, 7.5, 8.0, 8.5, 9.0, 9.5,\\ 10.0, 10.5, 11.0, 12.0]$ eV due to
the energy difference for the uphill process.  A total of 160000
trajectories for each initial condition were run with impact
parameters $b \in [0.0-14.0]$ a$_{0}$. Overall, 20790481
state-to-state rates were determined for the reverse reaction.\\

\noindent
For training the NN model, 12 diatomic properties were used as input
features: internal energy, vibrational energy, vibrational quantum
number, rotational energy, rotational quantum number, angular momentum
of the diatom, relative translational energy, relative velocity,
turning periods of the diatom, rotational barrier height, and
vibrational time period of the diatom.\cite{MM.sts:2019} Together with
the STS cross sections, this constituted the input to train the
updated forward (STS2025) and reverse NN-based models required for the
present work. From the energy dependent STS cross sections, $(\sigma_{v, j
  \rightarrow v', j'}(E_t))$ for given $(v, j, E_{t})$, the respective
kinetic STS reaction rate dictionary was built using the predicted NN
cross-section according to $k_{v, j \rightarrow v', j'}(E_t) = v_{\rm
  rel} \times \sigma_{v, j \rightarrow v', j'}(E_t)$, where
$v_{\rm{rel}}= \sqrt{\frac{8k_BT}{\pi \mu}}$ and $T = 10000$ K .\\

\noindent
In addition to the atom-exchange reactions, it was found that the
dissociation channels needed to be included for completeness in the
coarse grained modeling\cite{panesi:2022}. Consequently, further QCT
simulations for the two dissociation channels
N$_{2}(X^{1}\Sigma_{g}^{+}$) + O($^{3}$P) $\rightarrow$ N($^4$S) +
N($^4$S) + O($^3$P) and NO($X^2 \Pi$) + N($^4$S) $\rightarrow$
N($^4$S) + N($^4$S) + O($^3$P) were carried out. For the N+NO and
O+N$_2$ dissociation $\rightarrow$ N + N + O reactions, 50000
trajectories for each initial $(v,j)$ condition (8733 and 6329 for
N$_2$ and NO, respectively) were run using PES$_{\rm B}$. For each
$(v,j)$ initial condition, the collision energy was sampled from a
Maxwell-Boltzmann distribution at $T = 10000$ K. The initial distance
between the reactants was 20 a$_0$ with a maximum impact parameter
$b_{\rm max} = 14$ a$_0$.  The atoms were labeled as N$_{\rm A}$,
N$_{\rm B}$, and O so that (in)elastic and atom exchange
reactions can be distinguished. The number of dissociative
trajectories was counted independently for each initial $(v,j)$
state.\\

\noindent
For PES$_{\rm M}$, previous work\cite{panesi:2022} determined a total
of 12'742'744 STS kinetic rates for the NO($v,j$) + N
$\leftrightarrow$ N$_{2}$($v',j'$) + O reaction based on QCT
simulations. The same trajectories were also used to extract the
relevant information for full dissociation to atomic products which
constitutes the University of Illinois state-to-state model
(STS-UI).\\

\subsection{Master Equation Analysis}
Master equation simulations were carried out for an isothermal heat
bath condition to study the influence of the forward and reverse
kinetics obtained from the two different PESs. For the forward and
reverse heterogeneous exchange processes, the set of master equations
reads
\begin{equation}
    \frac{dn_i}{dt} = \sum_m^{\text{N}_2}
    \dot{\omega}_{m,i}^{E,\text{N}_2},
    \label{eq:master_1}
\end{equation}
\begin{equation}
    \frac{dn_m}{dt} = - \sum_i^{\text{NO}}
    \dot{\omega}_{m,i}^{E,\text{N}_2},
    \label{eq:master_2}
\end{equation}
\begin{equation}
    \frac{dn_{\text{O}}}{dt} = - \sum_i^{\text{NO}}
    \sum_m^{\text{N}_2} \dot{\omega}_{m,i}^{E,\text{N}_2},
    \label{eq:master_3}
\end{equation}
\begin{equation}
    \frac{dn_{\text{N}}}{dt} = \sum_i^{\text{NO}} \sum_m^{\text{N}_2}
    \dot{\omega}_{m,i}^{E,\text{N}_2},
    \label{eq:master_4}
\end{equation}
\begin{equation}
    \dot{\omega}_{m,i}^{E,\text{N}_2} = k_{m \rightarrow
      i}^{E,\text{N}_2} \big|_X n_m n_{\text{O}} - k_{i \rightarrow
      m}^{E,\text{NO}} n_i n_{\text{N}}.
    \label{eq:master_5}
\end{equation}
where $n_Y$ is the number density of the $Y$ species/level, $t$
denotes time, and $i$ and $m$ are multi-indices that refer to the NO
and $\text{N}_2$ rovibrational states,
respectively. $\dot{\omega}_{m,i}^{E,\text{N}_2}$ is the net mass
production rate due to the forward and reverse heterogeneous exchange
reactions (superscript $E$), NO($X^2 \Pi$) + N($^4$S)
$\leftrightarrow$ N$_{2}(X^{1}\Sigma_{g}^{+}$) + O($^{3}$P). $k_{i
  \rightarrow m}^{E,\text{NO}}$ denotes the state-to-state forward
exchange reaction rate. The reverse reaction rate, $k_{m \rightarrow
  i}^{E,\text{N}_2} \big|_X$, can be determined by employing either
the micro-reversibility (MR) (i.e., $k_{m \rightarrow
  i}^{E,\text{N}_2} \big|_{\text{MR}} = k_{i \rightarrow
  m}^{E,\text{NO}} / K_{i,m}^E$ where $K_{i,m}^E$ denotes the
equilibrium constant) or through direct calculation (DR) by means of
QCT simulations or evaluating the NN-trained models to yield $k_{m
  \rightarrow i}^{E,\text{N}_2} \big|_{\text{DR}}$. If
microreversibility is used, $K_{i,m}^E$ was determined from the
species' partition functions.\cite{panesi:2022}\\

\noindent
The master equations (Eqs. \ref{eq:master_1} to \ref{eq:master_4})
were then numerically integrated using \textsc{plato} (PLAsmas in
Thermodynamic nOnequilibrium) \cite{munafo:2020}, an object-oriented
library for nonequilibrium plasmas developed within the Center for
Hypersonics and Entry Systems Studies (CHESS) at the University of
Illinois at Urbana-Champaign. In the present study, the state-to-state
master equations in Eqs. \ref{eq:master_1} to \ref{eq:master_4} are
solved to obtain time evolution of the microscopic population
distributions, $n_i$ and $n_m$.\\

\noindent
The sums in Eqs. \ref{eq:master_1} and \ref{eq:master_2} over the
molecular internal states define the macroscopic reaction rate
coefficients, $k^{E,\text{N}_2}$ and $k^{E,\text{NO}}$, for the
forward and reverse heterogeneous exchange reactions. In this case,
the set of master equations governs the species' macroscopic
concentrations (i.e., $n_{\text{NO}}$, $n_{\text{N}_2}$,
$n_{\text{N}}$, and $n_\text{O}$) and $k^{E,\text{N}_2}$ and
$k^{E,\text{NO}}$ are defined using the Arrhenius parameters in Table
\ref{tab:tab1}. In addition to the chemical rates, a model is required
to describe the internal (\emph{i.e.} rovibrational) energy relaxation
of the molecular species, $\text{N}_2$ and NO. In the present study,
the concept of the conventional two-temperature (2-T) model
\cite{park1993review} is adopted for simplicity, resulting in that the
classical Landau-Teller formulation\cite{millikan:1963} was
employed:
\begin{equation}
    \frac{dE_{v}}{dt} = \frac{E_{v,\text{N}_2}(T) -
      E_{v,\text{N}_2}(T_v)}{\tau_{VT,\text{N}_2+\text{O}}} +
    \frac{E_{v,\text{NO}}(T) -
      E_{v,\text{NO}}(T_v)}{\tau_{VT,\text{NO}+\text{N}}},
    \label{eq:LT}
\end{equation}
where $E_v(T)$ and $E_v(T_v)$ are the average vibrational energy in
the equilibrium and the non-equilibrium states,
respectively. $E_{v,s}$ denotes the average vibrational energy of
species $s$. The vibrational-translational (VT) relaxation times of
the collision pairs considered in the present study,
$\tau_{VT,\text{N}_2+\text{O}}$ and $\tau_{VT,\text{NO}+\text{N}}$,
were taken from the work of Park \cite{park1993review}. It is noted
that for the VT energy transfer the STS and Arrhenius treatments
differ. In the Arrhenius treatment $\tau_{VT}$ describes the
bound-bound transition of the same species (e.g. inelastic +
homogeneous exchange), for example, N$_2(i)$+O$\rightarrow$N$_2(k)$+O
instead of product of NO($m$)+N. Conversely, for the STS treatment,
the rovibrational-translational (RVT) transfer occurs all
simultaneously based on the STS rate coefficients, which might include
the heterogeneous exchange, unlike the Arrhenius treatment. In the
discussion section, a somewhat more balanced approach is discussed.\\

\noindent
To handle detailed balance for multiple competing reactions the
individual kinetic processes were balanced
separately.\cite{park1989nonequilibrium} Hence, for the forward
reaction N$_2$+O$\rightarrow$NO+N the reaction N$_2$+O$\leftarrow$NO+N
is used for detailed balance, whereas for dissociation
N$_2$+O$\rightarrow$N+N+O the reverse reaction
N$_2$+O$\leftarrow$N+N+O is relevant for detailed balance. In other
words, each chemical kinetic process has a corresponding reverse
process, and the different reaction pathways remain independent of one
another, even when detailed balance is enforced.\\

\noindent
The master equation simulations were started from equal fractions of
NO and N unless otherwise stated. For the initial gas pressure and
internal temperature $p=1000$ Pa and $T=300$ K were used,
respectively, and the temperature of the surrounding heat bath was
10000 K.\\

\section{Results}

\subsection{The Two Potential Energy Surfaces}
\begin{figure}[ht]
    \centering \includegraphics[width=1.0\textwidth]{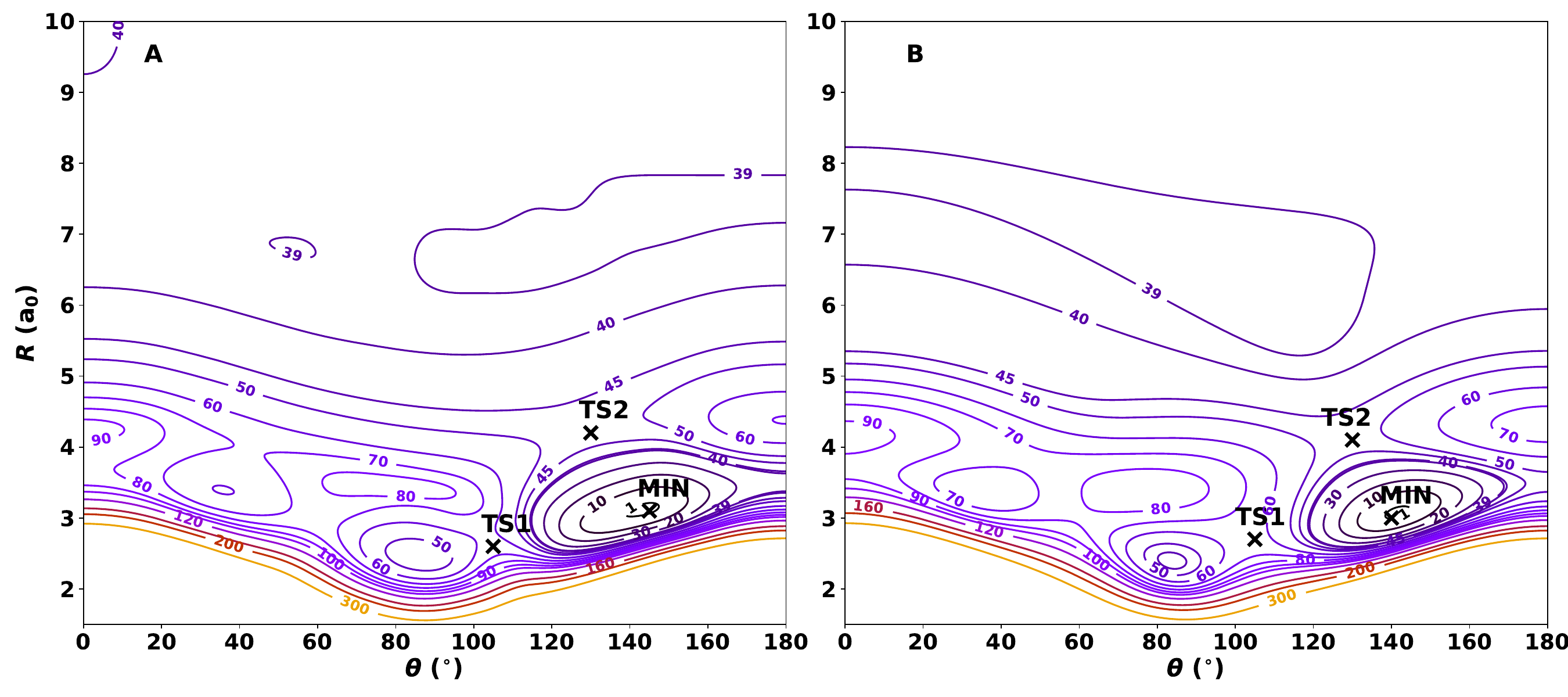}
    \caption{Representations of $V(R,\theta)$ for $r_{\rm NO}=2.30$
      a$_0$ in Jacobi coordinates for the N$_2$O RKHS-PES (panel A),
      and the PIP-PES (panel B). The energies (kcal/mol) are with
      respect to the global minimum of each PES. Energy-wise, TS1 and
      TS2 are at approximately 61.5 / 64.0 kcal/mol and 47.7 / 47.7
      kcal/mol relative to MIN in panels A and B, respectively. For
      both PESs isocontours are drawn at the same energies.}
    \label{fig:pess}
\end{figure}

The topography of the two $^3$A$'$ PESs used in the present work,
PES$_{\rm B}$ and PES$_{\rm M}$, is reported in Figure
\ref{fig:pess}. Both PESs are drawn at identical values of the
isocontours with the zero of energy taken as the global minimum of
each PES, respectively. The general shapes and critical points of
features of the two PESs are similar to each other. These include the
location of the global minimum at $[R=3.1 \,\mathrm{a}_0, \theta
  =144^{\circ}]$, the transition state at around $[R=2.7
  \,\mathrm{a}_0, \theta =103^{\circ}]$, and the distant transition
state at around $[R=4.2 \,\mathrm{a}_0, \theta =128^{\circ}]$. The
N$_2$O PIP-PES is generally steeper around the global minimum and the
two PESs differ in their anisotropy in the long-range part of the PES
($R > 6$ a$_0$). Still, given the different levels of electronic
structure theory and their representation, the similarities between
the PESs are rather striking.\\

\begin{figure}[h!]
    \centering \includegraphics[scale=0.7]{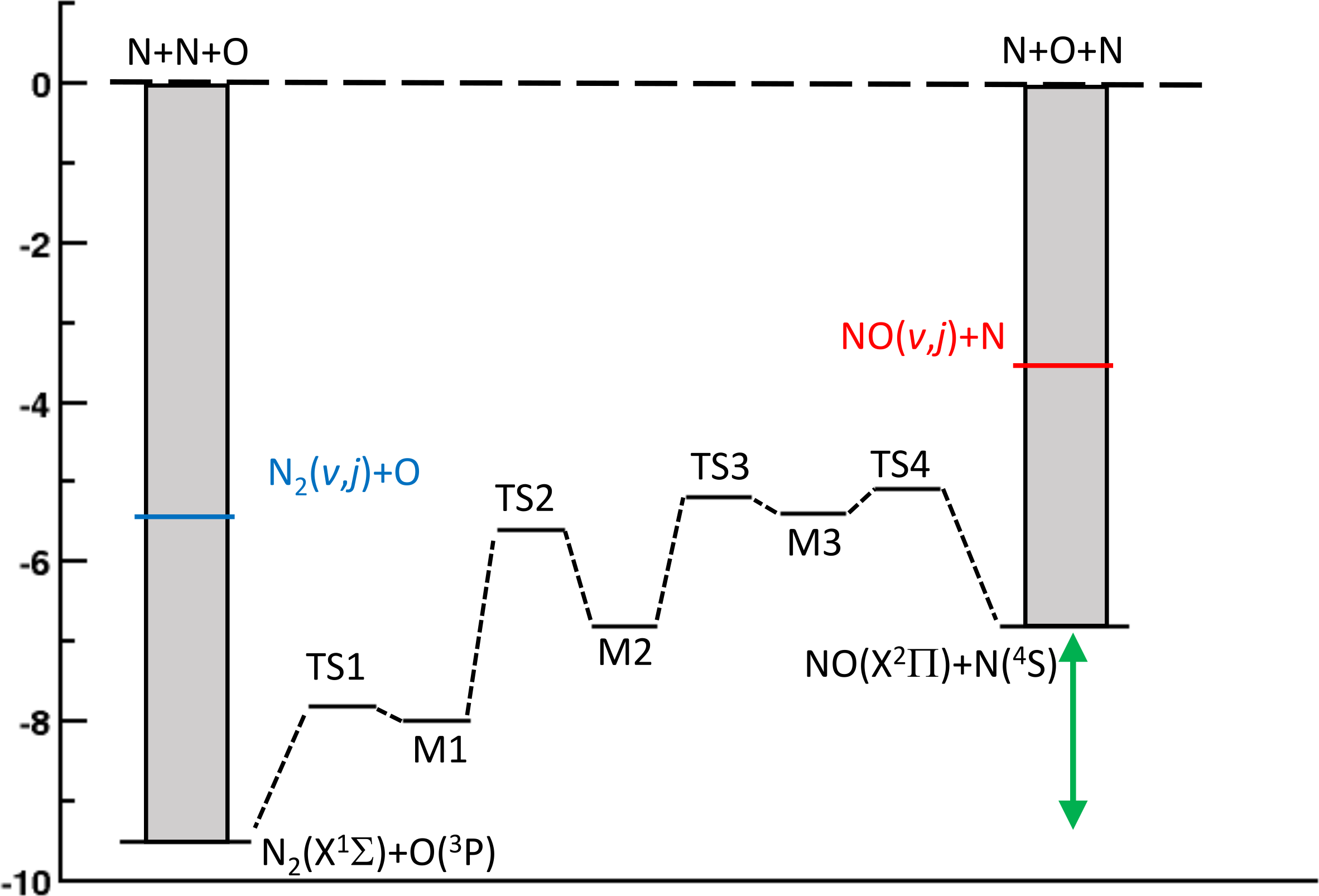}
    \caption{Energy level diagram. From right to left: NO(X$^2
      \Pi$)+N($^2$S), transition states (TS$i$) and local minima
      (M$j$), and N$_2$(X$^1 \Sigma$)+O($^3$P). The energetics of the
      states is from Ref.\cite{MM.n2o:2020}. The dissociation limit
      for N+N+O is indicated as the long-dashed line and the energies
      for the NO($v=13$,$j=119$) $E_{\rm int}=5.23$ eV; NO($E_{\rm
        int}=[0.12-9.80]$ eV) and N$_2$($v=16$,$j=135$) $E_{\rm
        int}=7.84$ eV; N$_2$($E_{\rm int}=[0.14-14.33]$ eV) states are
      given for reference in red and blue. The shaded areas indicate
      the 6329 and 8733 states for NO and N$_2$, respectively.}
    \label{fig:schematic}
\end{figure}

\noindent
For the forward (downhill, see Figure \ref{fig:schematic}) reaction
NO(X$^2 \Pi$) + N($^4$S) $\rightarrow$ N$_2$(X$^1 \Sigma_{\rm g}^{+}$)
+ O($^{3}$P) on the $^3$A$'$ PES two pathways were
reported.\cite{lin:2016,MM.n2o:2020} The first one amounts to N-attack
at the oxygen-side of NO, compression of the NON angle, formation of
N$_2$ and ejection of the oxygen atom, see Figure
\ref{fig:schematic}. An alternative pathway, not shown in Figure
\ref{fig:schematic}, starts at NO(X$^2 \Pi$) + N($^4$S) and directly
leads to minimum M1 over a single, lower-lying, transition state and
involves a collision of the incoming nitrogen atom with the
nitrogen-side of NO. TS4 is 32.9 kcal/mol above the entrance channel
for PES$_{\rm B}$ which compares with 37.3 kcal/mol for PES$_{\rm
  M}$. For the second pathway the single TS is 9.6 kcal/mol and 10.5
kcal/mol above the NO($X^2 \Pi$) + N($^4$S) entrance channel for
PES$_{\rm B}$ and PES$_{\rm M}$. PES$_{\rm B}$ has a 73.1 kcal/mol
energy difference between NO + N and N$_2$ + O, whereas for PES$_{\rm
  M}$ this value is 75.8 kcal/mol.\\

\subsection{The NN-based State-to-State Models}
\begin{figure}[h!]
    \centering \includegraphics[scale=0.4]{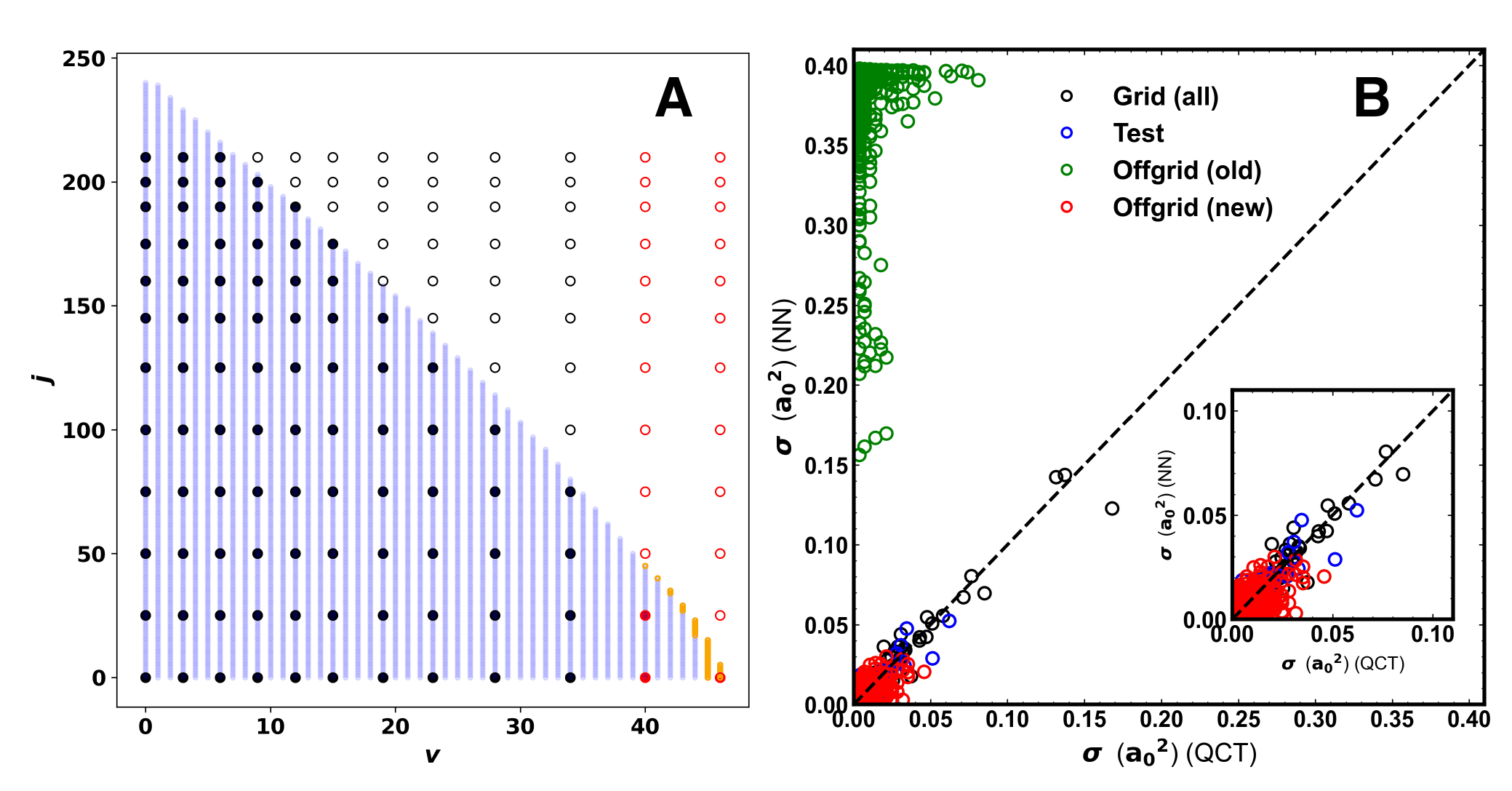}
    \caption{Panel A: The NO$(v,j)$ states from which QCT simulations
      were run for training STS2019 (black circles, $v_{\rm NO}\in
      [0,34]$ and $j_{\rm NO} \in [0,210]$) and additions for
      retraining STS2025 for the forward (downhill) reaction NO(X$^2
      \Pi$) + N($^4$S) $\rightarrow$ N$_2$ (X $^{1}\Sigma_{\rm
        g}^{+}$) + O($^{3}$P).  Black: NO$(v,j)$ grid used in the
      STS2019 model inside and outside the NO$(v,j)$ accessible state
      space are solid and open circles, respectively; Blue: accessible
      NO$(v,j)$ states; Orange: NO$(v,j)$ states for which STS2019
      predictions $\sigma_{\rm pred}^{\rm STS2019}$ disagree with QCT
      simulations for high-$v$ ($v=[40-47]$ and $j=[0-50]$). Red:
      additional grid to improve performance. The available
      state-space for NO(X$^2 \Pi$) includes $v \in [0,47]$ and $j \in
      [0,241]$. Panel B: Cross-section correlation between reference
      (QCT) and STS evaluated (NN) data. On-grid (black, STS2025);
      test grid (blue, STS2025); off-grid (green, STS2019); red
      off-grid (red, STS2025). This chart confirms that
      near-dissociation states need to be included explicitly.}
    \label{fig:sts}
\end{figure}

Computing complete and converged dictionaries for state-to-state rates
even for atom+diatom reactions is a daunting task. This motivated the
development of machine learning-based approaches (statistical models)
that require only a fraction of explicitly QCT-determined STS-rates
and predict the remaining rates from a trained neural
network.\cite{MM.sts:2019} Evaluating the trained model on the
$^3$A$'$ ground state of the NO($X^2 \Pi$) + N($^4$S) $\rightarrow$
N$_{2}$(X$^{1}\Sigma_{g}^{+}$) + O($^{3}$P) reaction for all
accessible state NO$(v \leq 46, j \leq 240)$ pointed to some
deficiencies of the STS2019 model.\cite{MM.sts:2019} The distribution
of all state-to-state cross sections (Figure \ref{sifig:stsrates2025})
using STS2019 (green trace in Figure \ref{fig:sts}) features a
non-negligible fraction $(10^{-5})$ of unusually large cross
sections. This prompted the explicit validation of such cross sections
compared with results from explicit QCT simulations. For this, $1.6
\times 10^5$ trajectories were run for $v=40,47$ for all $j \in [0,
  25, 50, 75, 100, 125, 145, 160, 175, 190, 200, 210]$ and $E_{\rm
  col}=0.86$ eV ($\approx 10000$ K) each initial condition at
$T=10000$ K. Upon analysis it was found that unreliable predictions
invariably concerned high $v-$states with $v_{\rm NO} \geq 35$,
i.e. near the NO dissociation threshold, see red circles in Figure
\ref{fig:sts}A. Consequently, the STS2019 model needed to be extended
with STS rates starting from high $v_{\rm NO}-$quantum number and
subsequent retraining of the NN.\\

\noindent
The additional QCT simulation required to improve and retrain the
STS2019 model yielded an additional 1261 non-zero STS cross sections
for the NO + N $\rightarrow$ N$_2$ + O reaction which were added to
the training set. After retraining the NN, model STS2025 was found to
be considerably improved, see the black and blue circles in Figure
\ref{fig:sts}B. Furthermore, the distribution of cross sections does
not contain unusually large values, red trace in Figure
\ref{sifig:stsrates2025}. It is also relevant to note that the
distribution of cross sections from STS2025 and those determined from
QCT simulations using PES$_{\rm M}$ are largely consistent, see Figure
\ref{sifig:stsrates2025}.\\

\noindent
The distribution of these STS cross sections (blue) and those from
models STS2019 (green) and STS2025 (red) is reported in Figure
\ref{sifig:stsrates2025}. For STS2019 the fraction of large STS cross
sections around $5 \times 10^{-12}$ a$_0^2$ is clearly visible. On the other
hand the range of the retrained STS2025 rates is closer to the STS
cross sections from simulations using PES$_{\rm M}$ whereas the shapes
of the blue and red distributions differ to some extent. This is
understandable as PES$_{\rm B}$ and PES$_{\rm M}$ differ in their
topography and the number of bound states they support and the number
of QCT simulations that were used for STS2019/STS2025 and for STS-UI
differ as well.\\

\noindent
The performance of the new STS2025 model for the reverse reaction is
shown in Figure \ref{sifig:sts-uphill}. For a test set the model
preforms satisfactorily for cross sections in the range [0-0.09]
$a_{0}^{2}$. Further improvement of the model, for example through
hyperparameter optimization, is in principle possible but was not
attempted in the context of the present study.\\

\subsection{Assessing Microreversibility from Arrhenius and STS-Based Rates}
In the following, the chemical evolution of the NO($X^2 \Pi$) +
N($^4$S) $\leftrightarrow$ N$_{2}(X^{1}\Sigma_{g}^{+}$) + O($^{3}$P)
system will be evaluated using different models and
approximations. PLATO determines the time evolution of the chemical
composition of the system given initial populations of the species
involved and either a thermal rate expression $k(T)$ such as an
Arrhenius fit or a dictionary of STS rates from a machine learned
model (STS2025) or from explicit QCT simulations (STS-UI). The
Arrhenius parameters (Arr$_{\rm B}$ and Arr$_{\rm M}$) from QCT
simulations using the two PESs (PES$_{\rm B}$ and PES$_{\rm M}$) are
reported in Table \ref{tab:tab1}.\\

\noindent
When running the PLATO simulations, the rates for the reverse
(``uphill'') reaction N$_{2}(X^{1}\Sigma_{g}^{+}$) + O($^{3}$P)
$\rightarrow$ NO($X^2 \Pi$) + N($^4$S) can either be determined from
assuming detailed balance using the available forward rates for
NO($X^2 \Pi$) + N($^4$S) $\rightarrow$ N$_{2}(X^{1}\Sigma_{g}^{+}$) +
O($^{3}$P), or the reverse rates need to be evaluated explicitly. This
can be accomplished from either using the STS2025 NN-based models (for
PES$_{\rm B}$) or from explicit QCT simulations (for PES$_{\rm M}$).\\ 

\begin{figure}[h!]
    \centering \includegraphics[scale=0.7]{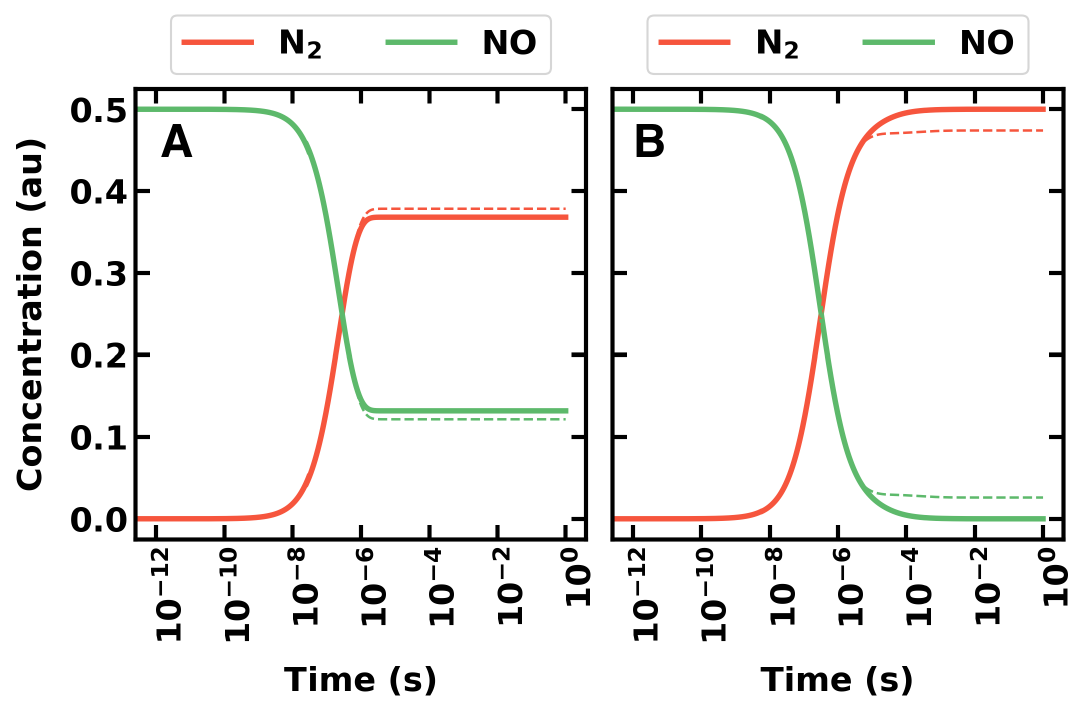}
    \caption{Concentration of N$_2$ (red) and NO (green) species as a
      function time for generating N$_2$ (NO + N$\rightarrow$N$_2$ +
      O, forward). All PLATO simulations use the $^3$A$'$ PES$_{\rm
        B}$ and the initial populations were [NO]$(t=0) = 0.5$ and
      [N$_2$]$(t=0) = 0$. Panel A: Arr$_{\rm B}$ parameters fitted to
      QCT simulations (see SI of Ref\cite{MM.n2o:2020}) for forward
      and MR for reverse reaction, i.e. using $[k^{\rm A}_{\rm f},
        k^{\rm A,MR}_{\rm r}]$ (dashed line); explicit forward and
      reverse rates, i.e. $[k^{\rm A}_{\rm f}, k^{\rm A}_{\rm r}]$
      (solid line). For further analysis including Arr$_{\rm B}$
      reverse and MR for the forward reaction, see Figure
      \ref{sifig:arrh.mrsi}. Panel B: Rates from STS2025 $[k^{\rm
          STS}_{\rm f}, k^{\rm STS,MR}_{\rm r}]$ (dashed) and $[k^{\rm
          STS}_{\rm f}, k^{\rm STS}_{\rm r}]$ (solid). Both, Arrhenius
      and STS do not converge to the correct equilibrium composition
      but to different degrees. To obtain the correct equilibrium
      composition one needs to include full dissociation for both
      models.}
    \label{fig:arr.sts}
\end{figure}

\noindent
First, the assumption of microreversibility for the reverse reaction
is tested for PES$_{\rm B}$. For this, PLATO simulations were carried
out using Arr$_{\rm B}$ parameters for the forward reaction (subscript
``f'') $k^{\rm A}_{\rm f}$ with reverse rates $k^{\rm A,MR}_{\rm r}$
from assuming microreversibility (dashed lines in Figure
\ref{fig:arr.sts}A). Using Arrhenius rates $[k^{\rm A}_{\rm f}$,
  $k^{\rm A}_{\rm r}]$ for the forward and reverse reaction (solid
lines in Figure \ref{fig:arr.sts}A) leads only to insignificant
changes. Hence, the results reported in Figure \ref{fig:arr.sts}A
indicate that the final N$_2-$concentration does not depend on whether
reverse rates are determined from assuming microreversibility or an
explicit Arrhenius rate is used. Also, the ``ignition point'' (time
$t_i$ at which [${\rm N}_2 (t_i)] = [{\rm N}_2(t \rightarrow
  \infty)]/2$) is $t_i = 0.27 \times 10^{-6}$ s when assuming
microreversibility compared with $t_i = 0.28 \times 10^{-6}$ s using
explicit rates for forward and reverse reactions. Results using
Arr$_{\rm M}$ parameters from PES$_{\rm M}$ are reported in Figure
\ref{sifig:arrh.mrsi} and confirm that PLATO simulations using
explicit reverse rates and those assuming microreversibility yield
essentially identical equilibrium concentrations and ignition
points.\\

\begin{table}[h!]
\centering
\begin{tabular}[t]{c|c|c|c|c}
  \hline
  \hline
  Reaction& $A$ & $n$ & $E_{\rm a}$ & Reference\\
  \hline
  NO+N$\rightarrow$N$_{2}$+O & $2.47 \times 10^{-12}$ & 0.4 & 8312 & Unibas\cite{MM.n2o:2020} \\
  NO+N$\rightarrow$N$_{2}$+O & $1.43 \times 10^{-13}$ & 0.8 & 6276 & UI\cite{panesi:2022} \\
  \hline
  N$_{2}$+O$\rightarrow$NO+N & $1.39 \times 10^{-10}$ & 0.1 & 47180 & Unibas \cite{MM.n2o:2020} \\
  N$_{2}$+O$\rightarrow$NO+N & $3.50 \times 10^{-11}$ & 0.4 & 48596 & UI\cite{panesi:2022} \\
  \hline
  \hline
\end{tabular}
\caption{Arrhenius parameters for thermal rates for the $^3$A$'$ state
  for PES$_{\rm B}$ and PES$_{\rm M}$ for the forward (NO+N) and
  reverse (N$_2$+O) reactions, respectively. Units are [$A$]=cm$^{3}$
  molecule$^{-1}$s$^{-1}$, [$E_{\rm a}$]= K and $n$ is unit-less.}
\label{tab:tab1}
\end{table}

\noindent
Next, the STS2025 model with and without assuming microreversibility
for the reverse rate - dashed and solid lines in Figure
\ref{fig:arr.sts}B - was used to follow the species' kinetics. The
equilibrium species' concentrations only differ by 0.03 units whether
or not explicit reverse rates were employed. Also, the ignition point
for both types of simulations are at $t_i = 0.31 \times 10^{-6}$
s. This is an increase by $0.04 \times 10^{-6}$ s compared with the
simulations based on Arrhenius-rates. The equilibrium product
concentrations between Arrhenius- and STS2025-based simulations
differ, however, by about 20 \%.\\

\noindent
In summary, the species' kinetics and ignition times $t_i$ do not
differ whether Arrhenius- or STS-based rates are employed but the
equilibrium amount of product generated differs by $\sim 20$ \%. This
point will be further discussed below.\\

\subsection{The Influence of the Underlying PES}
The two PESs - PES$_{\rm B}$ and PES$_{\rm M}$ - considered in the
present work support different numbers of bound rovibrational
states. This is due to the fact that a) the two PESs were determined
at different levels of theory (MRCI+Q/aug-cc-pVTZ for PES$_{\rm B}$
and MRCI/maug-cc-pVTZ with dynamical scaling (DSEC) for PES$_{\rm M}$,
respectively) and b) the reference energies were represented as a RKHS
and PIPs, respectively.\cite{MM.n2o:2020,lin:2016} Therefore, the
influence of the PES used in the QCT-simulations, i.e. PES$_{\rm B}$
vs. PES$_{\rm M}$, was probed next.\\

\begin{figure}[h!]
    \centering
    \includegraphics[scale=0.8]{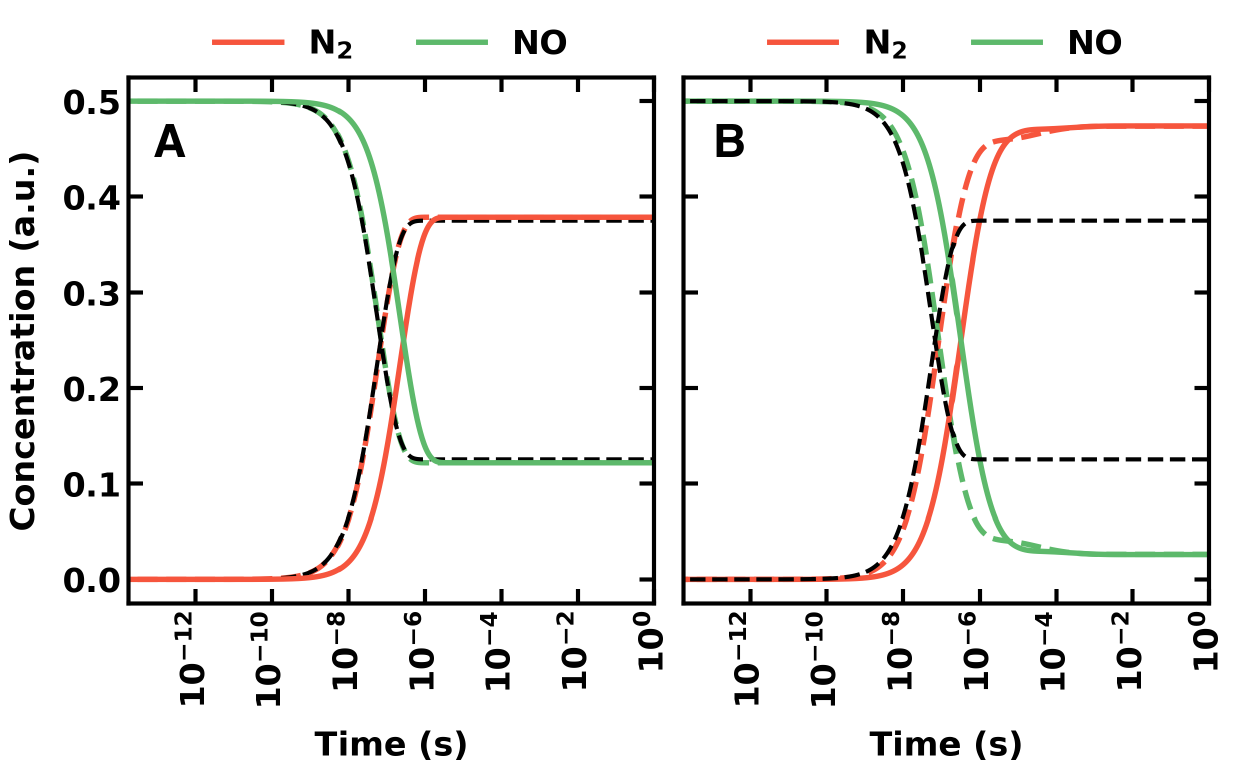}
    \caption{{\bf Effect of using reference data from two different
        PESs and MR-reverse rates} The initial populations in the
      PLATO simulations are [NO]$(t=0) = 0.5$ and [N$_2$]$(t=0) = 0$.
      Population of N$_{2}$ (red) and NO (green) as a function of
      time. Panel A assuming MR for the reverse reaction for PES$_{\rm
        B}$ $[k^{\rm A,Basel}_{\rm f}, k^{\rm A,Basel,MR}_{\rm r}]$
      (solid) and PES$_{\rm M}$ $[k^{\rm A,Minn}_{\rm f}, k^{\rm
          A,Minn,MR}_{\rm r}]$ (dashed). Panel B using state-to-state
      rates for the two PESs: $[k^{\rm STS2025}_{\rm f}, k^{\rm
          STS2025,MR}_{\rm r}]$ (solid) and $[k^{\rm STS-UI}_{\rm f},
        k^{\rm STS-UI,MR}_{\rm r}]$ (dashed). The black dashed lines
      in both panels represent the group-reconstructed approach, see
      Discussion.}
    \label{fig:arrh.mr_2_comp}
\end{figure}

\noindent
Figure \ref{fig:arrh.mr_2_comp}A compares the concentration profiles
from Arrhenius-rates for the forward reaction with reverse rates based
on assuming MR. Using PES$_{\rm B}$ the Arrhenius-rates Arr$_{\rm B}$
yield the same limiting concentration for $[{\rm N}_2(t \rightarrow
  \infty)] \sim 0.4$ (solid line) as does Arr$_{\rm M}$ determined
from running QCT simulations on PES$_{\rm M}$ (dashed line). The
ignition point from Arr$_{\rm M}$ occurs at $t_i = 0.72 \times
10^{-7}$ s compared with $t_i = 0.28 \times 10^{-6}$ s from
simulations using Arr$_{\rm B}$. When using the STS information from
QCT simulations on the two PESs (rates from STS2025 or STS-UI,
respectively) the limiting concentrations converge to $[{\rm N}_2(t
  \rightarrow \infty)] \sim 0.5$, see Figure
\ref{fig:arrh.mr_2_comp}B. Again, and consistent with using Arrhenius
rates, the ignition point from STS-UI is about 1 order of magnitude
earlier than that from using STS2025.\\

\begin{figure}[h!]
    \centering
    \includegraphics[scale=0.8]{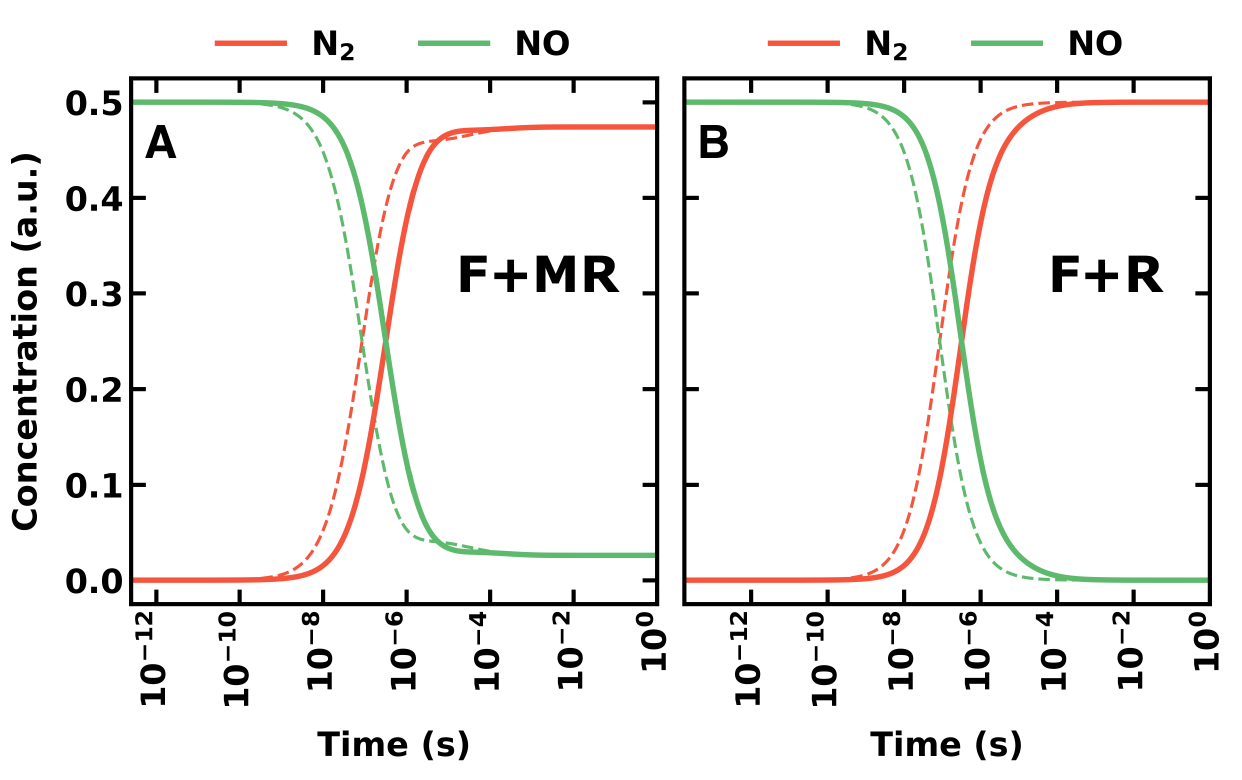}
    \caption{{\bf Effect of using reference data from two different
        PESs using STS rates} Temporal evolution of state-to-state
      derived mole fraction concentration at $T = 10000$ K. Panel A:
      Reverse rates from assuming microreversibility using $[k^{\rm
          STS2025}_{\rm f}, k^{\rm STS2025,MR}_{\rm r}]$ (solid) and
      $[k^{\rm STS-UI}_{\rm f}, k^{\rm STS-UI,MR}_{\rm r}]$
      (dashed). Panel B: Using explicit reverse rates $[k^{\rm
          STS2025}_{\rm f}, k^{\rm STS2025}_{\rm r}]$ (solid) and
      $[k^{\rm STS-UI}_{\rm f}, k^{\rm STS-UI}_{\rm r}]$
      (dashed). Panel A corresponds to Figure
      \ref{fig:arrh.mr_2_comp}B; here shown for direct comparison.}
    \label{fig:sts.bas.ill}
\end{figure}

\noindent
Finally, Figure \ref{fig:sts.bas.ill} compares the kinetics from using
MR-reverse rates (panel A) with those using explicit STS-reverse rates
(panel B) based on simulations with STS2025 and STS-UI. The data in
Figures \ref{fig:sts.bas.ill}A and \ref{fig:arrh.mr_2_comp}B is
identical but reported again for direct comparison. Figure
\ref{fig:sts.bas.ill}B shows that with explicit rates for the reverse
reaction the kinetics for both, STS2025 and STS-UI, yields
asymptotically $[{\rm N_2}] = 0.5$ on the $10^{-4}$s time scale. This
compares with $[{\rm N_2}] = 0.48$ if reverse rates are obtained from
assuming MR, see Figure \ref{fig:sts.bas.ill}A. The ignition points
when using STS2025 and STS-UI also remain unchanged whether reverse
rates are from assuming microreversibility (panel A) or explicitly
given from the STS dictionaries (panel B).\\

\noindent
In summary, the species' kinetics is surprisingly insensitive to
whether input data for the coarse grained simulations was obtained
from PES$_{\rm B}$ or PES$_{\rm M}$, except for a slightly different
ignition point. This pertains to both, whether or not
microreversibility for the reverse rates was assumed. Also, the
results from using Arrhenius-parameters instead of STS-information are
consistent and provide additional reassurance as to the robustness of
the findings independent on which of the two PESs was used.\\

\subsection{Kinetics Including the Dissociation Channels}
Up to this point the results from PLATO simulations using STS-data
generated from PES$_{\rm B}$ and PES$_{\rm M}$ are surprisingly
similar, see Figures \ref{fig:arrh.mr_2_comp} and
\ref{fig:sts.bas.ill}. However, modeling of reactive flow without
accounting for the possibility of the reaction products to fully
decompose into atomic fragments at temperatures $T=10000$ K or above
is not entirely realistic. This is reflected, for example, in the
equilibrium product concentrations, see Figures \ref{fig:arr.sts} to
\ref{fig:sts.bas.ill}: in all cases the diatomic species'
concentrations converge to finite values at the end of the
simulation. Given that all simulations
were carried out at 10000 K and in an isothermal heat bath, all
molecules should dissociate in thermodynamic equilibrium. This,
however, is not possible without including the dissociation
channel. Therefore, PLATO simulations were also carried out by
including fragmentation N$_{2}({\rm X}^{1}\Sigma_{g}^{+}$) +
O($^{3}$P) $\rightarrow$ N($^4$S) + N($^4$S) + O($^3$P) and NO(${\rm
  X}^2 \Pi$) + N($^4$S) $\rightarrow$ N($^4$S) + N($^4$S) + O($^3$P),
see Figure \ref{fig:schematic}. For PES$_{\rm B}$ the required initial
state-dependent dissociation rates were determined from explicit QCT
simulations. For PES$_{\rm M}$ the same procedure was followed.\\

\begin{figure}[h!]
     \centering \includegraphics[scale=1.0]{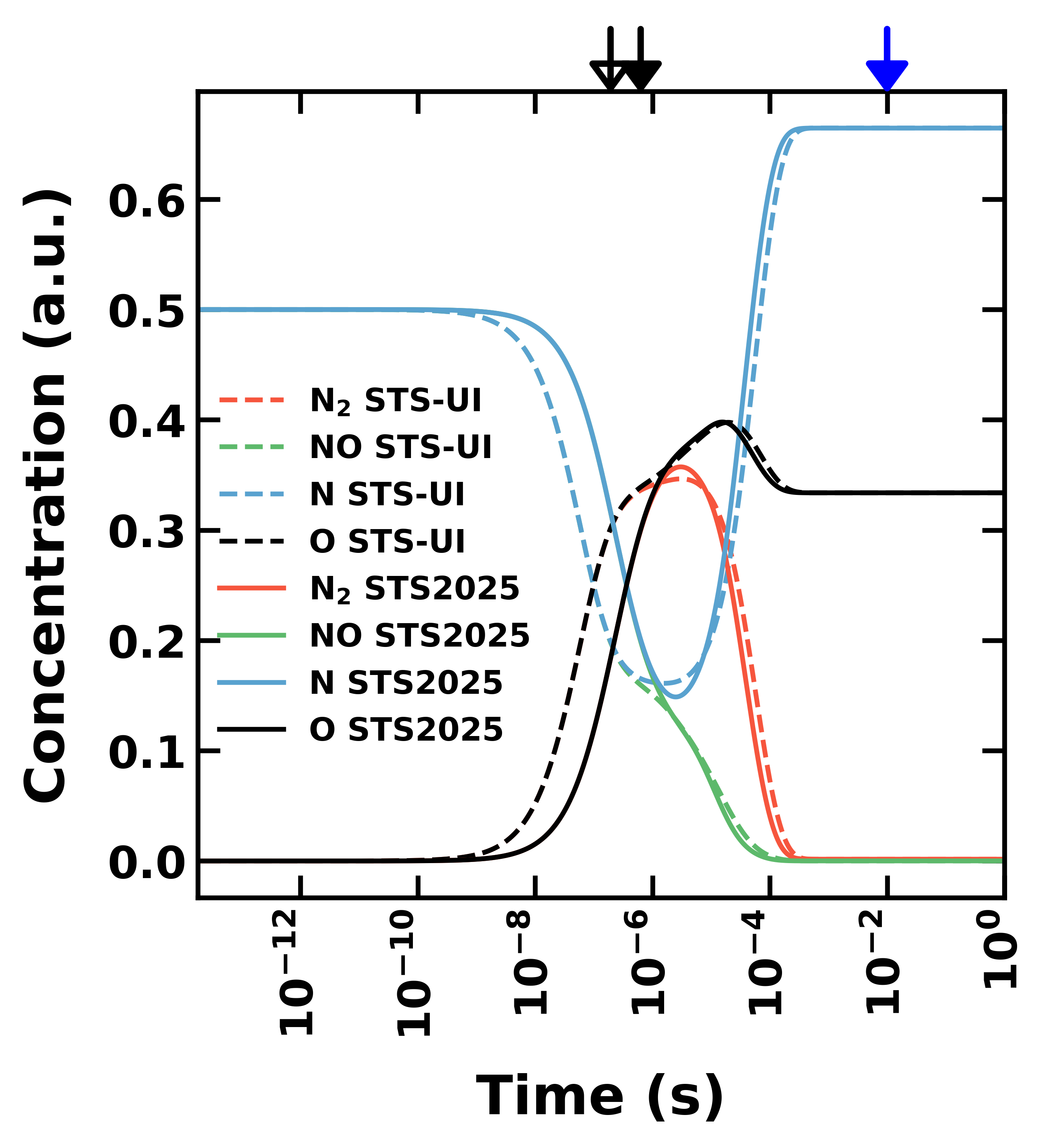}
     \caption{Effect of including the dissociation channel: Time
       evolution of state-to-state-derived mole fraction concentration
       at $T = 10000$ K. Forward rates from STS and MR-reverse rates
       on the $^3$A$'$ PES. Data for Panesi\cite{panesi:2022} using
       QCT-derived rates (STS-UI, dashed line) and NN-derived rates
       (STS2025, solid lines). The difference in concentrations for $t
       < 10^{-6}$ s is attributed to differences in the forward rates
       from the PESs. The behaviour for $t > 10^{-6}$ s is dominated
       by the dissociation kinetics. Both treatments yield comparable
       equilibrium distributions. The downward arrows indicate ignition points from 
        PES$_{\rm B}$ (solid black, $0.20 \times 10^{-6}$ s), PES$_{\rm M}$ (open black, $0.62 \times 10^{-6}$ s), and the stationary equilibrium concentration (solid blue).}
     \label{fig:compare}
\end{figure}

\noindent
Figure \ref{fig:compare} reports the species concentrations within
isothermal heat bath conditions (see Methods) as a function of time
over 15 orders of magnitude, until all reagents' concentrations were
stationary, which occurs after $t \sim 10^{-2}$ s. Simulations using
information from PES$_{\rm B}$ are reported as solid lines and those
using PES$_{\rm M}$ as dashed lines. The overall behaviour of both
models is comparable over the entire time range. However, the ignition
point for PLATO simulations from PES$_{\rm B}$ is later by about half
an order or magnitude compared with PES$_{\rm M}$. In other words, the
generation of atomic oxygen (black) is faster with data from PES$_{\rm
  M}$. The equilibrium compositions reached (1/3 for [O] and 2/3 for
[N], as required) are virtually identical and the temporal evolution
of [N$_2$], [NO], [N], and [O] match rather closely given the
different origins of the STS-rates, the dissociation rates and
differences in constructing and representing the PESs underlying the
QCT simulations.\\

\section{Discussion and Outlook}
The present work investigated the species' kinetics for
[N,O]-containing systems from coarse-grained simulations based on
atomically resolved state-to-state input using two different PESs. The
necessary STS-data was either obtained from a trained NN or from
explicit QCT simulations. It should be noted that vibration-rotation
$(v,j)-$coupling was not included in the present STS model but has
been explicitly considered in the past.\cite{MM.std:2022} In addition,
a ``double coarse-grained'' approach was considered whereby Arrhenius
rates - fitted to reaction rates from QCT simulations using both PESs
- were used as input for the PLATO simulations. Finally, the
assumption of reverse rates from detailed balance was explicitly
tested. The results so far indicate that despite the rather different
origins of the two PESs and the way how the STS dictionary was
generated, the species' kinetics is remarkably similar.\\

\begin{figure}[h!]
    \centering
    \includegraphics[width=0.97\linewidth]{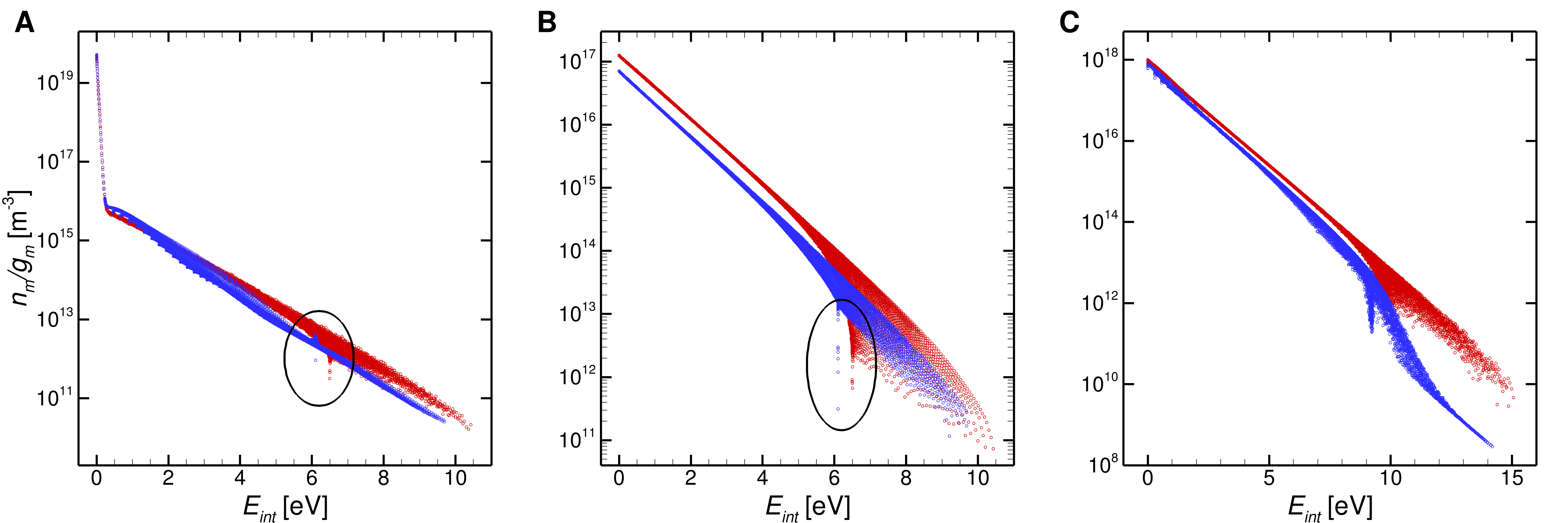}
    \caption{Population analysis at $T = 10000$ K with STS-information
      using STS2025 (blue) and STS-UI (red). Panel A: NO population at
      30\% mole fraction. Panel B: Quasi-steady state (QSS) population
      of NO. Panel C: QSS population of N$_2$. In panel A the features
      around 6 eV are due to a preferential sensitivity of states with
      $(v_{max}, j=0)$ for NO and N$_2$ for the exchange reaction. The
      pronounced decreases of $n_{m}/g_{m}$ in panels B and C arise
      from dissociation of NO and N$_2$, respectively. This is because
      the $(v_{max}, j=0)$ state has the largest dissociation rate,
      resulting in the severe depletion of $n_m$. See Figure
      \ref{sifig:Plot-Population_Separate} for separate panels for
      results from using STS2025 and STS-UI.}
    \label{fig:sts.pop}
\end{figure}

\noindent
As an additional point, thermochemical energy transfer depending on
the PES used was characterized. For this the microscopic states'
behavior during the nonequilibrium regime (i.e. for $t \leq 10^{-2}$
s, see Figure \ref{fig:compare}) was analyzed by directly comparing the microscopic rovibrational state populations predicted by the two sets of STS kinetics for selected
time intervals during the nonequilibrium regime of the chemical
kinetics. A common
approach\cite{liu2015general,panesi2013rovibrational,panesi:2022} for
analyzing the internal energy transfer and equilibration among the
rovibrational states is to compute the population distribution for the
different chemical species, here $\text{N}_2$ and NO. If the data is
represented as a semi-log-plot and at the thermochemical equilibrium
one therefore expects a linear relationship between the population and
the internal energy. Figure \ref{fig:sts.pop} shows the relationship
between the internal energy $e_m$ and the population of these levels
as quantified by $n_m / g_m$, where the degeneracy of the $m$-th state
is $g_{m} = g_{e}^{\rm N_{2}} g_{\rm nuc}^{\rm N_{2}} (2j(m) + 1)$;
$g_{e}$ and $g_{\rm nuc}$ correspond to the electronic (ground) and
the nuclear degeneracy, $j(m)$ is the rotational quantum number and
$n_{m}$ is the species population for the $m$-th ro-vibrational state
of N$_2$ (see Eqs. \ref{eq:master_1}-\ref{eq:master_5}). The
population distributions shown in Figure \ref{fig:sts.pop} were
extracted from the analysis presented in Figure \ref{fig:compare},
which includes both the N$_2$+O $\leftrightarrow$ NO+N, i.e. the
``Zeldovich kinetics'', and the dissociation kinetics. \\

\begin{figure}[h!]
    \centering
    \includegraphics[scale=0.11]{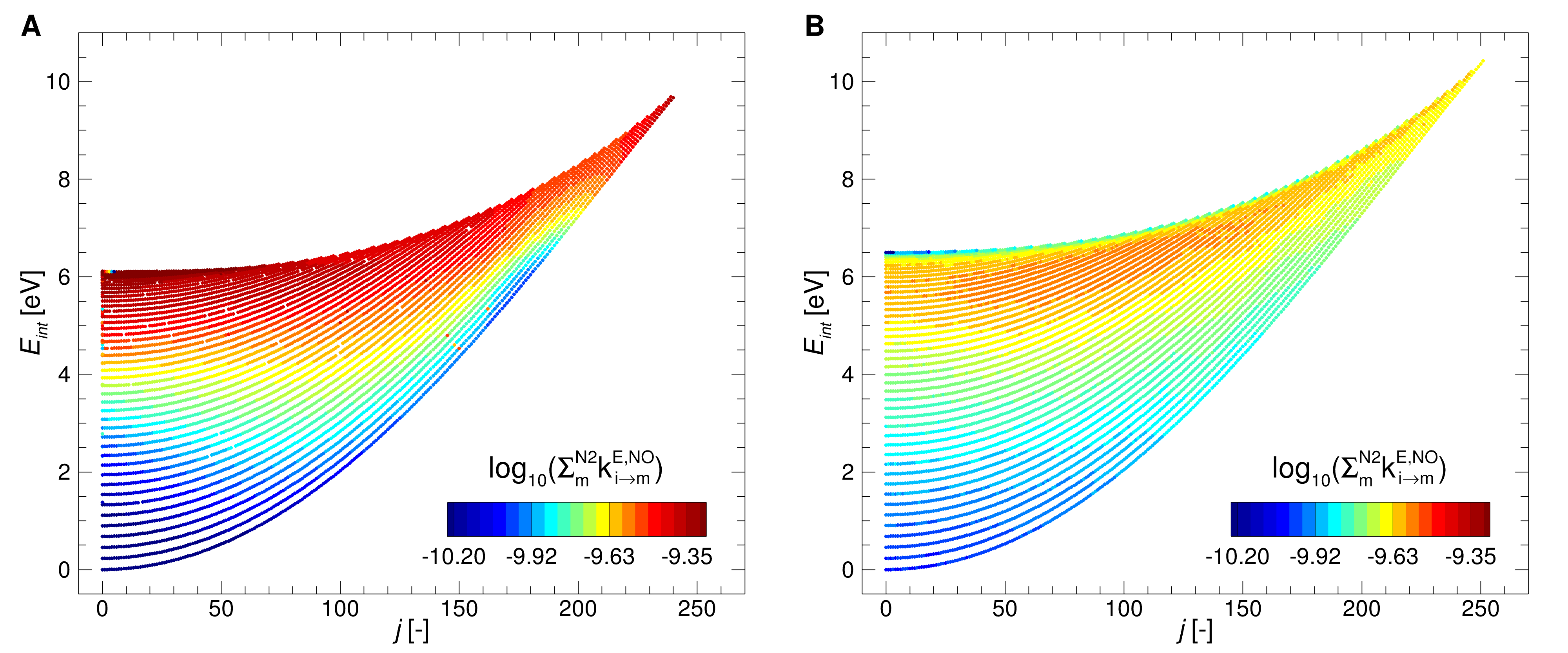}
    \caption{Distributions of the forward rates for NO$(i)$+N
      $\rightarrow$ $\text{N}_2(m)$+O at $T=10000$ K. The rates are
      summed over all possible product states and overlayed on the
      diatomic state space of NO. A comparison of the two PESs is
      made: STS2025 (left) and STS-UI (right).}
    \label{fig:sts.pop.image}
\end{figure}

\noindent
Figure \ref{fig:sts.pop}A shows the population distributions of NO at
30\% of its bulk species concentration as a result of the forward and
backward reactions between $\text{N}_2(m)$+O and NO$(i)$+N. This
corresponds to $t\approx 10^{-7}$ s, which was selected in particular
to analyze the impact of the exchange-driven process on the species
concentration before molecular dissociation takes control, see Figure
\ref{fig:compare}. The population distribution can be interpreted as a
degree of \emph{thermalization} by considering its level of scatter
("width") at a given energy. The narrower the distribution for given
energy, the faster thermalization occurs. As  Figure \ref{fig:sts.pop}A shows, STS2025 (blue) leads to more rapid
thermalization in the high-lying energy tails for internal energies
larger than 5 eV, indicated by the narrower width of the
distribution. On the other hand, between 1 eV and 4 eV, STS-UI yields
faster relaxation. This can be attributed to differences in the
microscopic forward reaction rates shown in Figure
\ref{fig:sts.pop.image}A and B which reports the microscopic rate
coefficients in the forward direction, $\text{NO}(i)+\text{N}
\rightarrow \text{N}_2(m)+\text{O}$. The net outgoing rates ending up
in internal states of $\text{N}_2$ (\emph{i.e.}  $\sum_m^{\text{N}_2}
k_{i \rightarrow m}^{E,\text{NO}}$) are relatively higher in the
high-$v$ region compared to low-$v$. For energies higher than 4 eV,
STS2025 features larger STS-probabilities, whereas in the energy range
1 eV to 4 eV, STS-UI predicts higher rates, resulting in the trend
shown in Figure \ref{fig:sts.pop}A in terms of the thermalization
during nonequilibrium relaxation. \\

\begin{figure}[h!]
    \centering \includegraphics[scale=0.11]{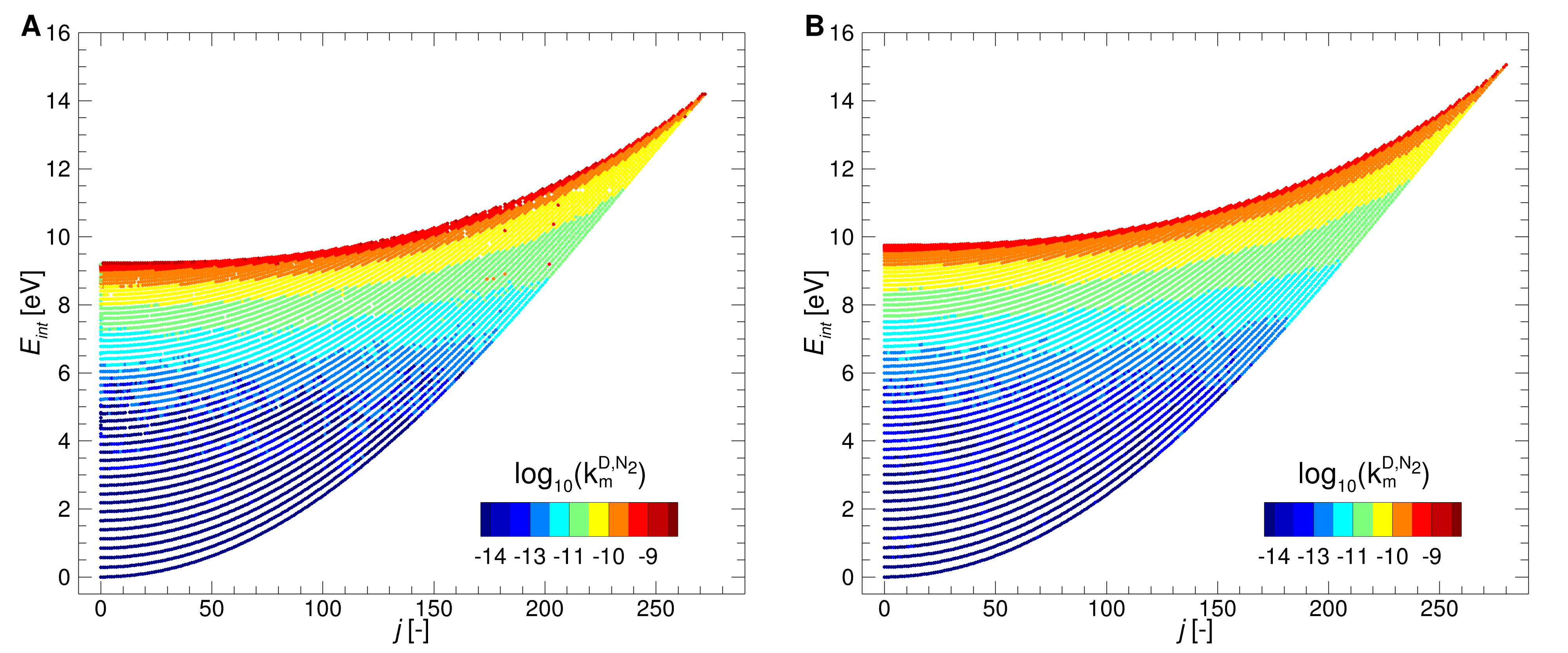}
    \caption{Distributions of the dissociation rates for
      $\text{N}_2$+O $\rightarrow$ N+N+O at $T=10000$ K. The rates are
      overlayed on the diatomic state space of $\text{N}_2$. A
      comparison of the two PESs is made: STS2025 (left) and STS-UI
      (right).}
    \label{fig:sts.pop.image_Diss}
\end{figure}

\noindent
Figures \ref{fig:sts.pop}B and C present the population distributions
for the quasi-steady state (QSS) regime of NO and $\text{N}_2$,
respectively. The QSS regime is characterized by the fact that the
dominant chemistry that occurs in this time interval is the
destruction of NO and N$_2$, respectively. The starting times for the
molecular QSS periods are correspondingly $t_{\text{QSS}}^{\text{NO}}
\approx 10^{-6}$ s and $t_{\text{QSS}}^{\text{N}_2} \approx 10^{-5}$
s, see Figure \ref{fig:compare}. During the QSS periods, it is
expected that the dissociation reactions dominate the kinetic
process. For NO, STS-UI shows a wider scatter of the quasi-bound
states along the $y$-axis, whereas STS2025 leads to faster
thermalization in the high-lying energy around 6 eV and beyond. For
N$_2$ (see Figure \ref{fig:sts.pop}C) the differences between STS2025
and STS-UI concern primarily the quasi-bound energy region, above 9.75
eV. This is mainly because STS2025 predicts higher dissociation rates
compared to STS-UI, as shown in Figure
\ref{fig:sts.pop.image_Diss}. For the high-lying energy levels near
the dissociation limit (\emph{i.e.}  $\sim 9.75$ eV), STS2025 features
higher dissociation rates for a given $\left( j, E_{int} \right)$
pair. This might be attributed to the difference in the potential
well-depth of N$_2$ between the two PESs. The shallower well-depth of
PES$_{\text{B}}$ on the N$_2$ side results in a more compact
probability distribution within the diatomic energy space. This leads
to the larger dissociation in the high-lying tail as shown in Figure
\ref{fig:sts.pop}C. \\

\noindent
Based on the QSS state populations shown in Figures \ref{fig:sts.pop}B
and C, macroscopic dissociation rate coefficients, $k_g^D$, of NO and
N$_2$ were calculated according to $k_g^D = \sum_i n_i k_i^D / \sum_i
n_i$.\cite{panesi2013rovibrational} Here, $k_i^D$ is the
state-specific dissociation rate coefficient for state $i$. The
expression for $k_g^D$ yields the population-weighted global
dissociation rate for a particular diatomic species. For
NO-dissociation via NO+N collisions, STS-UI and STS2025 yield $9.82
\times 10^{-13}$ $\text{cm}^3/\text{s}$ and $1.51 \times 10^{-12}$
$\text{cm}^3/\text{s}$ whereas for N$_2-$dissociation through N$_2$+O
collision, they are $4.54 \times 10^{-14}$ $\text{cm}^3/\text{s}$ and
$5.90 \times 10^{-14}$ $\text{cm}^3/\text{s}$, respectively. The rates
from using the two PESs are consistent with one another, in particular
given the differences in the topography of the two under- lying PESs
considered in the present work and the different QCT-simulations they
are based on.\\

\noindent
Although the Arrhenius and STS approaches lead to the same conclusions
for the two PESs, it is worth noting that the two methods yield
somewhat different final concentrations, see Figure \ref{fig:arr.sts},
if molecular dissociation is not considered. This is mainly attributed
to the inherent differences in the physical modeling between the two
methods: The Arrhenius approach uses the macroscopic reaction rate
coefficient to describe the rate of change of the species
concentration, while the STS method explicitly describes the change of
the individual energy levels in the full resolution. In addition, from
a multi-temperature modeling point-of-view, the Arrhenius approach
relies on a first-order approximation (\emph{i.e.} Landau-Teller
formula) to describe energy transfer among the Boltzmann energy
pool of states. The required relaxation time $\tau_{VT}$ defined in
Eq. (\ref{eq:LT}) is often proposed in the
literature\cite{millikan:1963} for the combined inelastic and
homogeneous exchange processes, rather than that for heterogeneous
exchange, which is of interest in the present study. Hence, some
differences between Arrhenius and STS-based PLATO simulations are
expected, see Figure \ref{fig:arr.sts}.\\

\noindent
To provide additional insight into this difference further analyses
were carried out. An Arrhenius-based approach only considers the bulk
species composition, whereas a full STS treatment explicitly follows
rovibrational state populations as a function of time. To assess the
impact of treating the chemical step at two different levels of
resolution (1 rate expression for Arrhenius vs. $\sim 10^8$ cross
sections and rates for STS) on the results of the master equation
modeling it is instructive to consider an "intermediate model". Such
an approach was originally developed for probing the dissociation
kinetics of the O+O$_2$ system\cite{Venturi2020} and the same strategy
was also applied to the forward heterogeneous exchange process for the
combined $^3 A'$ and $^3 A''$ system of [NNO].\cite{panesi:2022}
Related to this, recent studies showed that defining relaxation times
$\tau_{\rm VT}$ in a self-consistent manner within a multi-temperature
modeling framework can predict the non-equilibrium species
concentration and the aerothermal heat
loads\cite{kim2021modification,grover2025comparative} which is not
possible from a conventional two-temperature Park
model.\cite{park1993review} \\

\noindent
The "intermediate model" introduces group-reconstructed
rates\cite{Venturi2020,panesi:2022}
\begin{equation}
  \label{eq:group-recon}
    \tilde{k}_{i \rightarrow m}^{E,\text{NO}} = k_T^{E,\text{NO}}
    \frac{g_m \exp\left( -\frac{E_m}{k_B T}
      \right)}{\sum_m^{\text{N}_2} g_m \exp\left( -\frac{E_m}{k_B T}
      \right) }
\end{equation}
where $k_T^{E,\text{NO}}$ is the thermal rate from Table
\ref{tab:tab1}, "E" refers to the exchange reaction, and $g_m$ and
$E_m$ are the degeneracy and the rovibrational energy of the $m$-th
state of $\text{N}_2$, respectively. In other words, the rates
$\tilde{k}_{i \rightarrow m}^{E,\text{NO}}$ are obtained from
Boltzmann-weighting the thermal $k_T^{E,\text{NO}}$ at temperature
$T$,\cite{panesi:2022} and Eq. \ref{eq:group-recon} provides a
route to treat an Arrhenius-based model at the equivalent resolution
of the STS model. Using the rates $\tilde{k}_{i \rightarrow
  m}^{E,\text{NO}}$ in the PLATO simulations it is found (dashed black
lines in Figure \ref{fig:arrh.mr_2_comp}A) that the results agree with
the corresponding Arrhenius-based approach. Therefore, increasing the
dimensionality from the thermal to the rovibrational STS does not
appreciably change concentration profiles. On the other hand, compared
with a full STS-treatment reported in Figure
\ref{fig:arrh.mr_2_comp}B, the ignition point from the "intermediate
model" remains unaltered but the final concentration deviates. From
this it is concluded that the inherent non-equilibrium nature of the
rovibrational state dynamics which can not be captured by an
equilibrium Arrhenius-treatment ultimately leads to differences in the
final concentrations.\\

\noindent
In summary, coarse-grained simulations for the NO(X$^2 \Pi$) +
N($^4$S) $\leftrightarrow$ N$_2$(X$^1 \Sigma_{\rm g}^{+}$) +
O($^{3}$P) reaction using two different $^3$A$'$ PESs yield remarkably
similar $t-$dependent macroscopic behavior. This includes
population-weighted reaction rates and the species concentration
history. On the other hand, the microscopic view of the rovibrational
state populations presents distinct differences between the PESs, in
particular near the quasi-bound levels of NO and $\text{N}_2$\, as
shown in Figure \ref{fig:sts.pop} which, however, do not noticeably
change the results of the coarse grained simulations, see Figure
\ref{fig:compare}.  Coarse-grained simulations using Arrhenius- and
STS-based treatments for the rates leads to comparable ignition points
but equilibrium populations of the products differ by 20 \%. This is
most likely due to the non-equilibrium nature of the state-dynamics
which is correctly handled within a STS-treatment but not captured
when using (equilibrium) Arrhenius rates. The rather close agreement
of the species' kinetics irrespective of the PES used indicates that
the dependence and sensitivity of the final results are moderate and
largely irrelevant, at least for the system considered here.  \\

\section*{Acknowledgment}
Support by the Swiss National Science Foundation through grants
200021-117810, the NCCR MUST (to MM), and the University of Basel is
also acknowledged. Part of this work was supported by the United
States Department of the Air Force, which is gratefully acknowledged
(to MM). This work was partially supported by the National Research
Foundation of Korea through a grant RS-2025-00522088 funded by the
Ministry of Science and ICT (to SMJ). The authors thank Dr. A. Munafò
and Prof. M. Panesi for providing access to the PLATO physico-chemical
library.\\

\section*{Supporting Information}
The supporting information reports the initial conditions on the
ground state PES, individual photodissociating trajectories, the
crossing seams between the singlet PES, and wavefunctions for the
fundamentals on the two singlet PESs.

\section*{Data Availability Statement}
Data accompanying the present study is available at
\url{https://github.com/MMunibas/plato}. \\

\clearpage

\bibliography{bib}

\clearpage

\renewcommand{\thetable}{S\arabic{table}}
\renewcommand{\thefigure}{S\arabic{figure}}
\renewcommand{\thesection}{S\arabic{section}}
\renewcommand{\d}{\text{d}}
\setcounter{figure}{0}  
\setcounter{section}{0}  
\setcounter{table}{0}

\newpage

\noindent
{\bf SUPPORTING INFORMATION: Reaction Dynamics for the [NNO] System
  from State-Resolved and Coarse-Grained Models}

\begin{figure}
    \centering
    \includegraphics[scale=0.6]{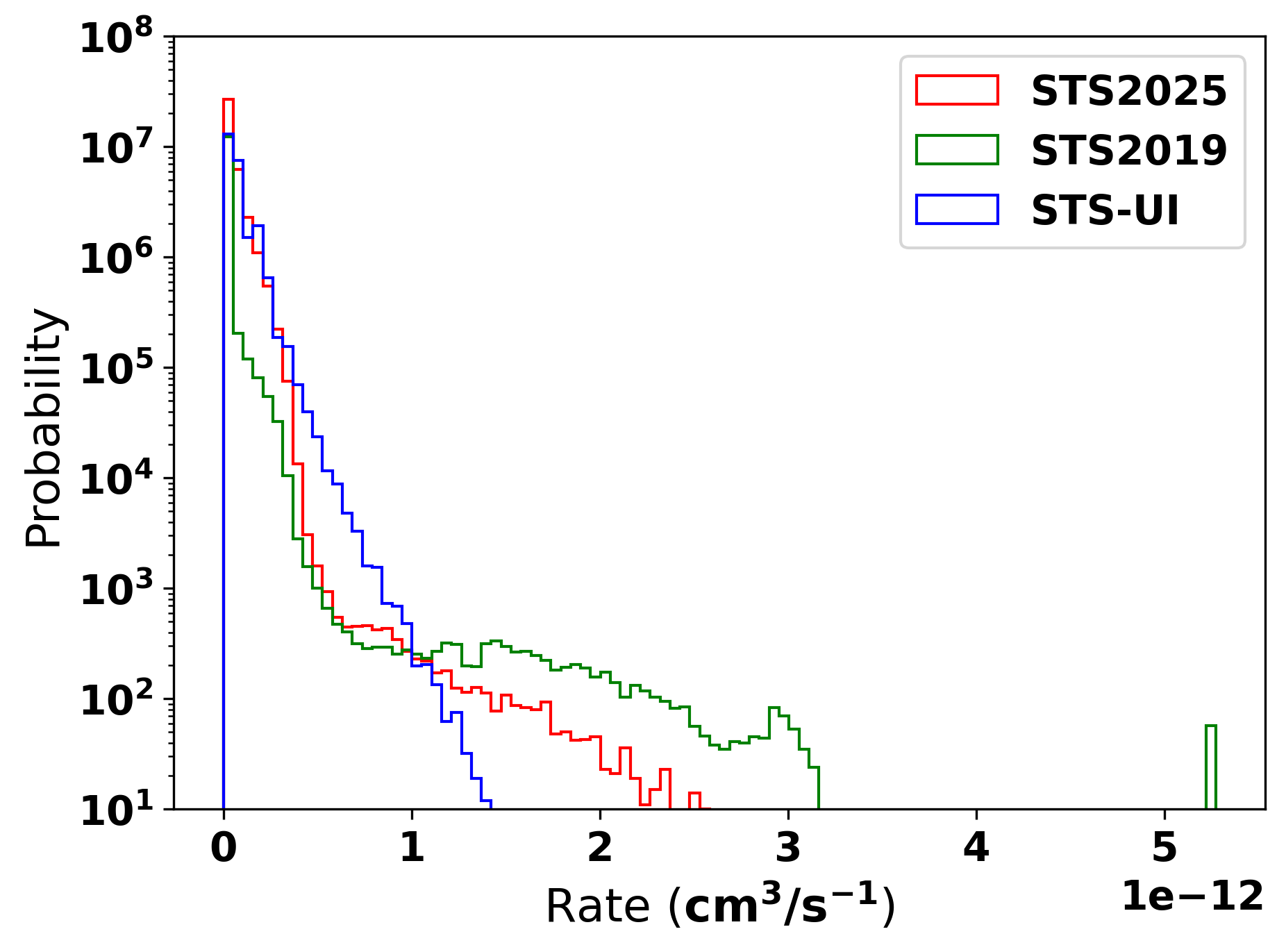}
    \caption{Rate (cm$^{3}/{\rm cm}^{-1}$) comparison for the forward
      NO$(v,j)$ + N $\rightarrow$ N$_{2}(v',j')$ + O reaction between
      STS-UI and STS2025. Rates $k_{v, j \rightarrow v', j'}(E_t) =
      v_{\rm rel} \times \sigma_{v, j \rightarrow v', j'}(E_t)$, where
      $v_{\rm rel} = \sqrt{\frac{8 k_B T}{\pi \mu}}$ were determined
      from cross sections evaluating STS2025.}
    \label{sifig:stsrates2025}
\end{figure}

\begin{figure}[h!]
    \centering \includegraphics[scale=0.9]{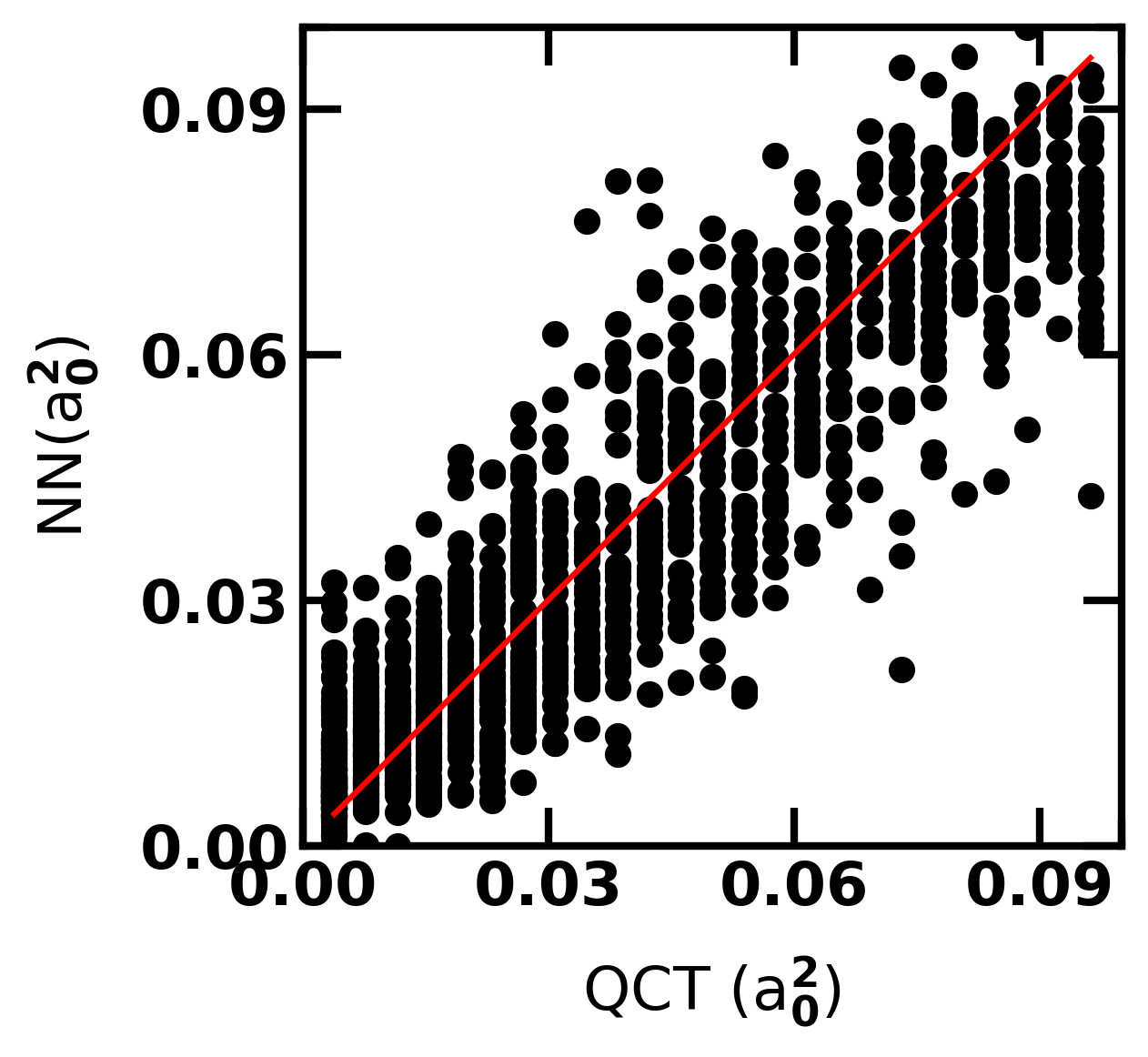}
    \caption{Performance of the STS2025 model for the reverse (uphill)
      reaction N$_{2}(X^{1}\Sigma_{g}^{+}$) + O($^{3}$P) $\rightarrow$
      NO($X^2 \Pi$) + N($^4$S) on the test set.}
    \label{sifig:sts-uphill}
\end{figure}

\begin{figure}[h!]
    \centering
    \includegraphics[scale=1.0]{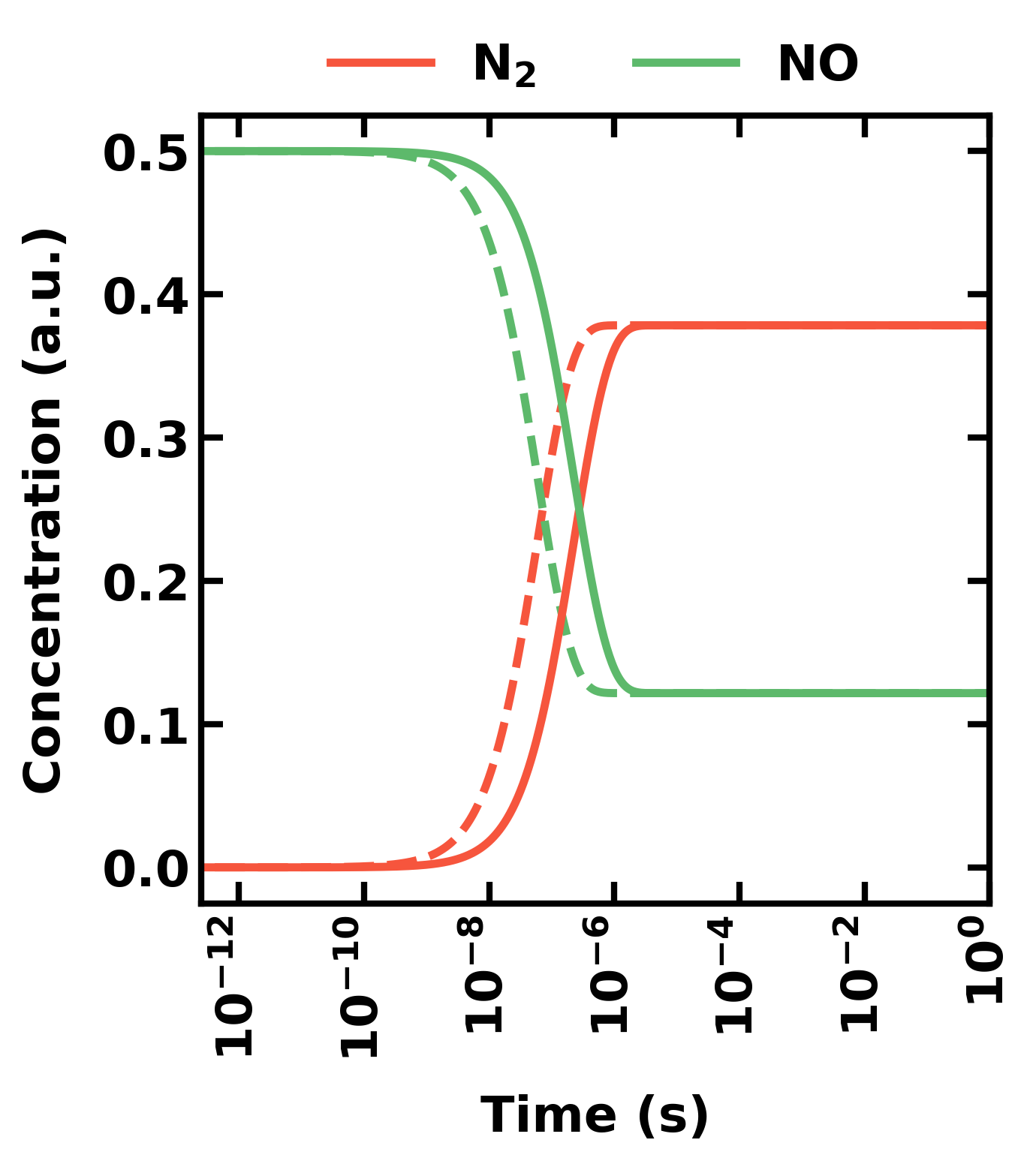}
    \caption{Population of N$_{2}$ (red) and NO (green) molecule as a
      function time for the NO + N$\rightarrow$N$_{2}$ + O (reverse)
      and N$_{2}$ + O $\rightarrow$ NO + N (forward) processes. The
      initial populations are [NO]$(t=0) = 0.5$ and [N$_2$]$(t=0) =
      0$.  Arrhenius for forward and microreversibility for reverse
      reaction, Basel data (solid line) and UIUC (dash line) are
      represented.  The Arrhenius parameters are fitted to the
      $^3$A$'$ results in the SI of Ref\cite{MM.n2o:2020} determined
      from QCT simulations using PES$_{\rm B}$.}
    \label{sifig:arrh.mrsi}
\end{figure}

\begin{figure}
    \centering \includegraphics[width=0.9\linewidth]{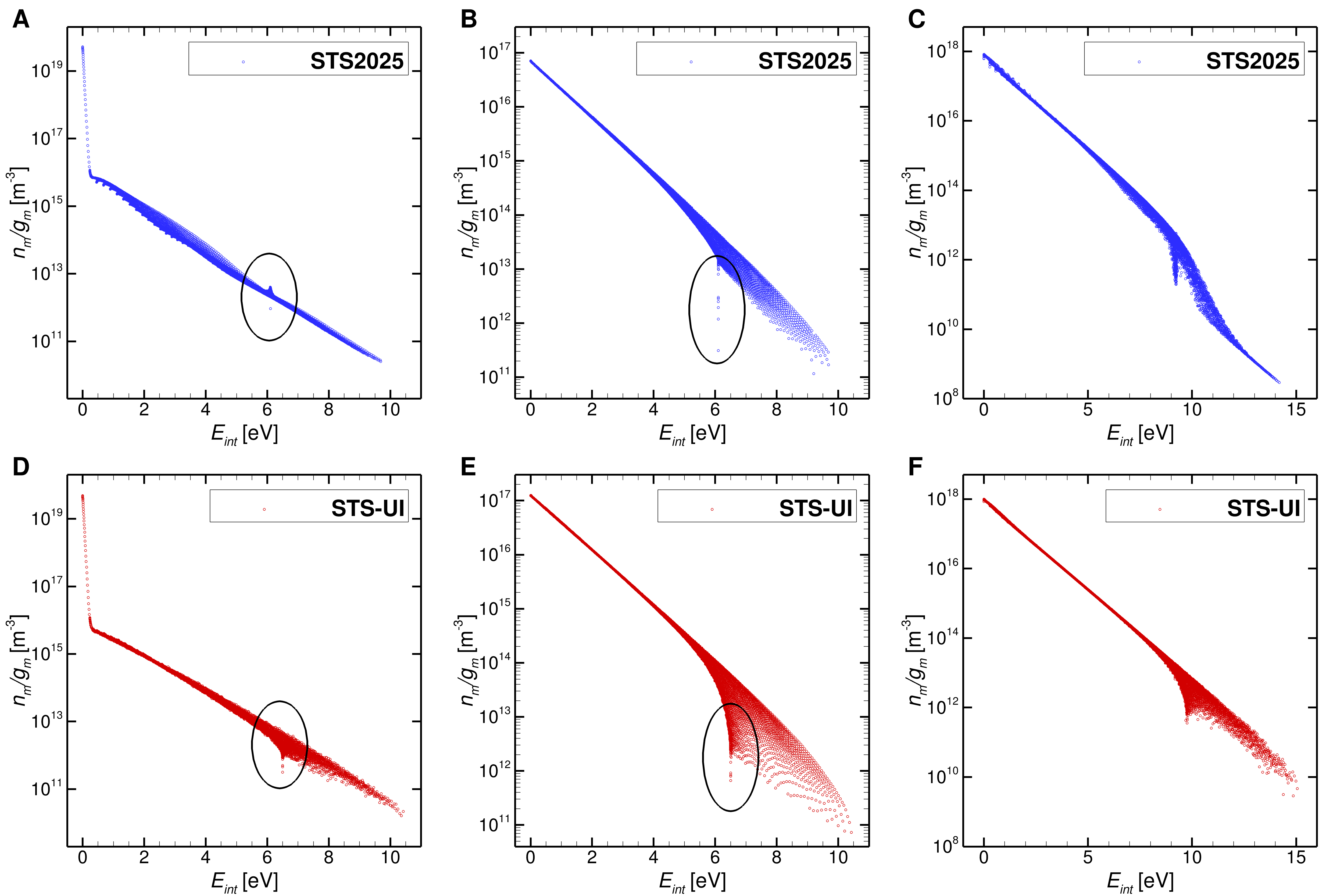}
    \caption{Population analysis at $T = 10000$ K with STS-information
      using STS2025 (top row, blue) and STS-UI (bottom row,
      red). Panels A/D: NO population at 30\% mole fraction. Panels
      B/E: Quasi-steady state (QSS) population of NO. Panels C/F: QSS
      population of N$_2$. In panels A/D the features around 6 eV are
      due to a preferential sensitivity of states with $(v_{max},
      j=0)$ for NO and N$_2$ for the exchange reaction. The pronounced
      decreases of $n_{m}/g_{m}$ in panels B/E and C/F arise from
      atomization of NO and N$_2$, respectively. This is because the
      $(v_{max}, j=0)$ state has the largest dissociation rate,
      resulting in the severe depletion of $n_m$. This figure is
      related to Figure \ref{fig:sts.pop} in the main manuscript.}
    \label{sifig:Plot-Population_Separate}
\end{figure}

\end{document}